\documentclass[a4paper,10pt]{JHEP3}
\usepackage{amsmath}
\usepackage{amssymb}
\usepackage{graphicx}
\usepackage{verbatim}
\usepackage{float}
\usepackage[small]{caption}
\usepackage{subfig}
\usepackage{booktabs}

\newcommand{\bra}[1]{\langle #1|}
\newcommand{\ket}[1]{|#1\rangle}

\def\Im{{\rm{Im}\,}}
\def\Rea{{\rm{Re}\,}}

\def\dblone{\hbox{$1\hskip -1.2pt\vrule depth 0pt height 1.6ex width 0.7pt\vrule depth 0pt height 0.3pt width 0.12em$}}
\newcommand{\be}{\begin{equation}}
\newcommand{\ee}{\end{equation}}
\newcommand{\bea}{\begin{eqnarray}}
\newcommand{\eea}{\end{eqnarray}}
\newcommand{\bal}{\begin{align}}

\newcommand{\enl}{\end{align}}

\newcommand{\ds}{{\sffamily DarkSUSY}}

\def\ov{\over  }

\title{Relic densities including Sommerfeld enhancements in the MSSM}
\author{Andrzej Hryczuk, Roberto Iengo, Piero Ullio \\ \\
SISSA and  INFN, Sezione di Trieste\\
via Bonomea 265, 34136 Trieste, Italy \\
E-mail: \email{hryczuk@sissa.it},\email{iengo@sissa.it},\email{ullio@sissa.it}}

\abstract{We have developed a general formalism to compute Sommerfeld enhancement (SE) factors for a multi-state system of fermions, in all possible spin configurations and with generic long-range interactions. We show how to include such SE effects in an accurate calculation of the thermal relic density for WIMP dark matter candidates. We apply the method to the MSSM and perform a numerical  study of  the relic abundance of neutralinos with arbitrary composition and including the SE due to the exchange of the $W$ and $Z$ bosons, photons
and Higgses. We find that the relic density can be suppressed by a factor of a few in a sizable region of the parameter space, mostly for Wino-like neutralino with mass of a few TeV, and up to an order of magnitude close to a resonance.
}


\begin{document}

\section{Introduction}

In the latest years a number of experiments have contributed to develop a deep understanding of the standard model of cosmology. Most
cosmological parameters have been measured with a level of precision hardly foreseeable a decade ago; among these,  there is the cold dark 
matter (CDM) contribution to the Universe energy density. E.g., a recent analysis, within the 6-parameter $\Lambda$CDM model, of the 7-year 
WMAP data~\cite{Jarosik:2010iu}, combined with the distance measurements from the baryon acoustic oscillations in the distribution of  
galaxies~\cite{Percival:2009xn} and the recent redetermination of the Hubble Constant with the Hubble Space Telescope~\cite{Riess:2009pu}, 
gives \cite{Komatsu:2010fb}:
\begin{equation}
 \Omega_{\rm{CDM}} h^2 = 0.1123\pm0.0035 \qquad {\rm{(68\%\ CL\ uncerinities)}},
\end{equation} 
where $\Omega$ indicates the ratio between mean density and critical density, and $h$ is the Hubble constant in units of $100\,{\rm km}\,{\rm s}^{-1}\,{\rm Mpc}^{-1}$.

The nature of the CDM term is still unknown; one of the most attractive scenarios is that dark matter is a thermal relic from the early Universe:
in this context, stable weakly interacting massive particles (WIMPs) are ideal CDM candidates, as their thermal relic abundance is naturally of the order of the 
measured one (for a recent review on WIMP DM see, e.g.,~\cite{DMbook}). Since the accuracy in the experimental determination of $\Omega_{\rm{CDM}}$ has reached 
the few per cent level, it is necessary to have equally accurate theoretical predictions for the relic abundance of WIMPs. Indeed, there may be a number of effects involved in 
this calculation, requiring a proper treatment and possibly making the final result differ even by orders of magnitude from the simple and naive estimate that
the relic density scales with the inverse of the total WIMP pair-annihilation cross section in the non-relativistic regime, and it takes the appropriate value when the 
annihilation rate is about $3\cdot 10^{-26}\;{\rm cm}^3\,{\rm s}^{-1}$. Such features include, e.g., threshold or resonance effects, co-annihilation 
effects (namely the interplay among several particles, nearly degenerate in mass, rather than a single WIMP) or three-body final states. 
Lately, some attention has been dedicated the case of WIMPs with annihilation cross sections strongly dependent on the velocity of the incoming pair. 
Such interest has been triggered by the fact that cross sections much larger the standard value seems to be needed to provide a DM positron 
source accounting for the rise in the positron fraction in the local cosmic-rays that has been measured by  PAMELA~\cite{Adriani:2008zr}: the idea is to 
explain such mismatch in terms of the mismatch between the velocity of the WIMPs at freeze-out in the early Universe, soon after entering the non-relativistic regime, 
i.e. for about $v/c\sim 0.3$, with the velocity of WIMPs in our Galactic halo, i.e. about  $v/c\sim 10^{-3}$, see, e.g.,~\cite{ArkaniHamed:2008qn}.  
A feature of this kind is predicted by to the so-called Sommerfeld enhancement~\cite{Sommerfeld}, namely the increase in the cross section occurring for highly 
non-relativistic particles in the presence of a long-range attractive force that deforms the wave function of the incoming pair with respect to the standard plane wave  
approximation. 

The Sommerfeld enhancement effect in the context of WIMP DM has been studied first in Refs.~\cite{Hisano:2002fk,Hisano:2004ds}; these authors considered, within
a supersymmetric extension of the standard model of elementary  particles, the case of pure Wino or pure Higgsino DM and showed that, when these states 
are very massive, at the TeV scale or heavier, the weak force can play the role of a long-range force since its carriers, the W and Z bosons, are much lighter. 
The impact on the relic density calculation for this case was discussed in Ref.~\cite{Hisano:2006nn}. Later, the Sommerfeld enhancement and its implication
on the relic density has been studied in the context of the minimal dark matter model~\cite{Cirelli:2007xd,Cirelli:2008jk}, still referring to standard interactions and very 
massive WIMPs.  More recently, in light of the PAMELA anomaly, several authors, including, e.g.,~\cite{ArkaniHamed:2008qn, Dent:2009bv, Zavala:2009mi, Slatyer:2009vg, Lattanzi:2008qa, Hannestad:2010zt, Feng:2010zp, Iminniyaz:2010hy, Mohanty:2010es, Arina:2010wv,Buckley:2009in,Feng:2009hw,MarchRussell:2008yu}, have considered 
a few aspects of the connection with WIMP DM, considering models for which a new dark force, carried by some yet-to-be-discovered light boson, provides a 
very large enhancement in the cross section. In our analysis we reconsider the case of supersymmetric DM, and discuss the effect in the general context of 
the minimal supersymmetric extension to the standard model (MSSM), with arbitrary Wino and Higgsino components in the lightest neutralino,\footnote{We 
will restrict to the case of negligible bino fraction since there is no Sommerfeld enhancement on trivial representation of $SU(2)_L$.} applying and 
developing the formalism introduced in Ref.~\cite{Iengo:2009ni} (the derivation of the Sommerfeld enhancements from field theory was discussed also in \cite{Cassel:2009wt}, see also \cite{Visinelli:2010vg}). Such formalism is more suitable to treat a realistic case of neutralino DM than the approach based 
on implementing a non-relativistic effective action followed by previous work on this subject \cite{Hisano:2002fk} (see also \cite{Cirelli:2007xd}). 

The new calculation that we present in this paper is then used to refine the computation of the thermal relic density of neutralinos. To solve
the corresponding set of Boltzmann equations we use the very accurate method given in the public available \ds\ package~\cite{Gondolo:2004sc}, appropriately modified 
to include Sommerfeld enhancement effects. Hence, we provide here a tool for very high precision estimates of relic abundances of neutralino DM in the 
MSSM, at a level comparable or better than the accuracy in the value for $\Omega_{\rm{CDM}} h^2$ from current and upcoming observations.
\par
The paper is organized as follows: In Section 2 we introduce the Sommerfeld enhancement and sketch the limits in which it may be relevant, while Section 3 discusses our approach to the case of $N$-coupled states. In Section 4 we detail the technique we use to implement the Sommerfeld effect in the relic density computation. Section 5 discusses our particle physics setup, while in Section 6 we present the results within this framework. Section 7 concludes.

\section{Sommerfeld enhancement}
\label{sec:somn}

We start with an short overview of the Sommerfeld effect itself and recap of the literature on the subject. The Sommerfeld enhancement is a non-relativistic effect resulting in correcting the cross section (in our case of interest the annihilation cross section) due to presence of some ``large distance force'' between the particles in the incoming state. It is generally described as an effect of distorting the initial wave function $\psi(\vec{r})$ of the incoming two-particle state by a non-relativistic potential. This potential is taken to be Yukawa or Coulomb, as the force arises due to exchange of massive or massless boson. The whole effect can then be encoded into the ratio between the wave function of the incoming, free particle (at $r\rightarrow \infty$) and the distorted one at the point of annihilation (at $r=0$). Thus one usually defines the enhancement factor being
\begin{equation}
 S=\frac{|\psi(\infty)|^2}{|\psi(0)|^2}\, ,
\end{equation}
which multiplies the cross section
\begin{equation}
 \sigma_{\rm{full}}=S\cdot \sigma_0\, .
\end{equation}
Many authors have discussed this effect introducing a new interaction, mediated by some scalar or vector boson $\phi$. If one assumes one species of annihilating particle $\chi$ with mass much larger than the mass of force carrier, $ m_\chi\gg m_\phi$, and a coupling of the order of the weak scale or larger, then one can get a large enhancement of the cross section. This can have a very important implications for indirect dark matter searches and relic density computations.
\par
The standard approach to estimate the enhancement is to introduce an interaction potential in the form:
\begin{equation}
 V=\alpha \frac{e^{-m_\phi r}}{r}\, ,
\end{equation}
and solve the Sch\"{o}rdinger equation for the incoming two-particle state to find the wave function distortion; there are however a few issues one should be careful about. First, the coefficient $\alpha$ in the potential depends on the nature of the incoming state and the possible type(s) of interaction. In particular, when the effect involves fermions, it is different if the interacting particles are Dirac or Majorana and whether they are identical or not; this is especially important when co-annihilations enter the computation of the relic density. 

To stress other delicate points, it is useful to recap first what are the conditions for the enhancement to be sizable. The Sommerfeld enhancement can be viewed as occurring due to forming a loosely bound state due to long-range interaction. In order to have such a bound state the characteristic Bohr energy of the interaction need to be larger than the kinetic energy. In the limit $m_\phi\rightarrow 0$, this gives a condition $\alpha^2 m_{\chi} \gtrsim m_\chi v^2$, i.e.: 
\begin{equation}
 v\lesssim \alpha\, .
\end{equation}
For a typical WIMP the coupling is of order $\alpha\sim 0.03$, so that there can be some sizable enhancement only long after freeze-out (which happens for $v\sim 0.3$). However, if there exists a slightly heavier state, then it may happen that just after freeze-out DM particles have enough energy to produce it nearly on-shell. At threshold these heavier states are produced with, roughly speaking, zero velocity. As we will see later on, if the mass splitting of the DM and the heavier state is small enough, this may give rise to important changes in the relic density.
\par
A second condition comes from the comparison of the range of the Yukawa potential with the Bohr radius. In order for the interaction to distort the wave function significantly, the range of the potential cannot be much smaller than the Bohr radius of the two-particle state, 
\begin{equation}
\label{longrange}
 \frac{1}{m_\phi}\gtrsim\frac{1}{\alpha m_{\chi}}\, .
\end{equation}
In case of very large enhancements, this condition needs to be even stronger, i.e. the range of potential has to be much larger than the Bohr radius. However, even in case of enhancements of order unity one should treat carefully the regime when $m_\phi\approx\alpha m_{\chi}$.
\par
When considering a system of two states with small mass splitting $\delta m$ interacting off-diagonally there is another important constraint. If $\delta m$ is significantly larger than the kinetic energy, it may seem that the heavier state cannot be produced, and hence there is no enhancement. However, if the potential is strong enough, there still may be an effect, coming from producing the heavier state at small distances, where the potential energy is large. Thus the condition reads:
\begin{equation}
 2\delta m \lesssim \alpha^2 m_{\chi}+\mathcal{E}\, ,
\end{equation}
meaning that the characteristic Bohr energy of the potential plus the kinetic energy $\mathcal{E}$ is large enough to produce the heavier state. Moreover, when dealing with multi-state systems, the picture of the Sommerfeld enhancement as an effect of a static, long range force is no longer applicable. This is because the exchange of $\phi$ leads to a momentum and energy transfer due to the mass splitting $\delta m$ and one may need to take into account terms of order $O(\delta m/m)$ modifying the interaction potential. 
\par
When dealing with a specific particle physics setup like, e.g., the MSSM, several such complications may intervene at the same time: a simple parametric description is not possible and for a proper estimate of the Sommerfeld effect and its impact on the relic density, a computation within a fully general formalism is needed. We introduce this in the next Section.

\section{Sommerfeld effect for $N$ coupled states}

We are interested in the general case with $N$ two-particle fermionic states coupled together. They interact with ``long-range forces'' due to the exchange of some boson $\phi$, denoting generically a vector, an axial vector or a scalar boson (in the MSSM those correspond to $Z^0$, $W^{\pm}$, $\gamma$, $H_{1,2}$ and $H^{\pm}$). The Sommerfeld enhancement corresponds to computing in the non-relativistic limit the sum of ladder diagrams as presented on the Fig. \ref{diag}a.
\FIGURE{\label{diag}
		 \includegraphics[scale=0.41,bb=71 422 1103 720]{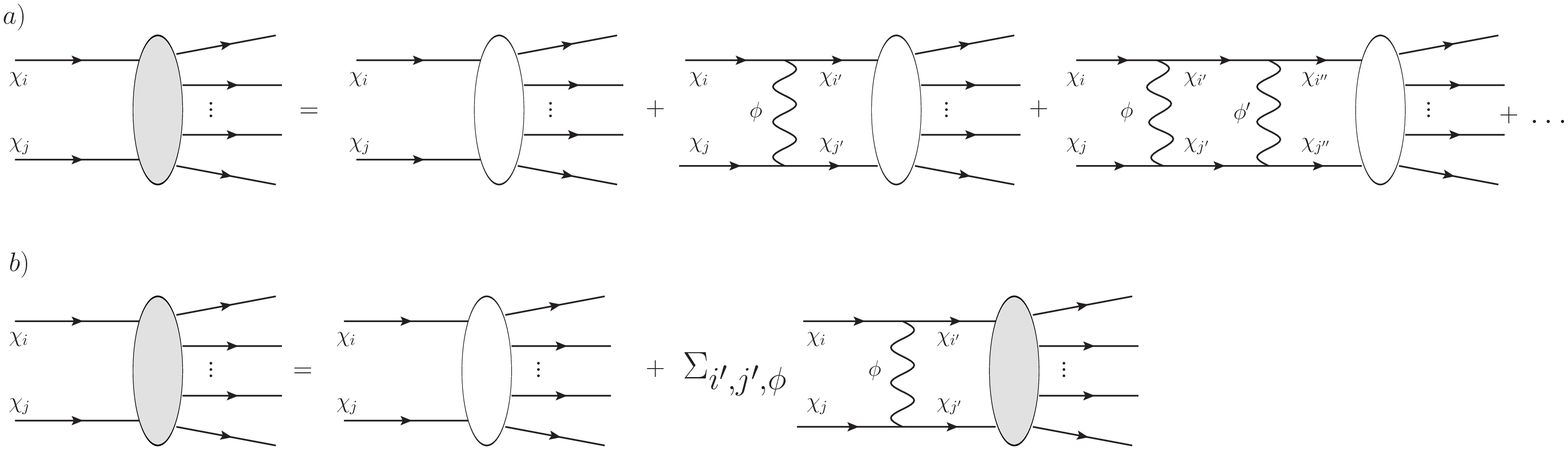}
		\caption{Ladder diagrams for the Sommerfeld enhancement. $a)$ incoming $\chi_i\chi_j$ particles interact with exchange of $\phi$'s (in general different), which can be a scalar, vector or axial vector bosons. In the ladder a virtual states $\chi_{i'}\chi_{j'}$ can be produced and the final annihilation proceed in the channel  which can be different than initial one. Filled blob represents full annihilation process with any number of SM particles in final state, while the empty one its tree level counterpart. $b)$ the same but written in a recursive form; sum is over all possible intermediate states and exchanged bosons.} }
The structure of those diagrams can be in general very complicated, since all states can appear in one diagram, through different interactions. The approach we follow to address the this problem is a generalization of the one developed in \cite{Iengo:2009ni}. We start from writing the recurrence relation for the annihilation amplitudes as presented pictorially on Fig. \ref{diag}b. Let's consider a process of the type
\begin{equation}
 \chi_a \chi_b\rightarrow \chi_i \chi_j \rightarrow \chi_i' \chi_j' \rightarrow \ldots \rightarrow \rm{SM\ final\ states\, ,}
\end{equation}
where the intermediate pairs $\chi_i\chi_j$ can be the same or different as the initial pair $\chi_a\chi_b$. The spin of the initial pair, which in the non-relativistic limit is a conserved quantity, can be in general either in the singlet ($S=0$) or triplet ($S=1$) state. For every possible $\chi_i\chi_j$ pair we get a recurrence relation for its annihilation amplitude. Denoting this amplitude by $A_{ij}$ and its tree level value by $A^0_{ij}$, one obtains in the non-relativistic limit \cite{Iengo:2009ni}:
\begin{equation}
\label{inteqn}
A_{ij}(p)=A^0_{ij}(p)-\sum_{i'j'\phi}N_{ij,i'j'}\frac{g_{ii'\phi} g_{j'j\phi}}{(2\pi)^3}\int \frac{d^3k}{(\vec{p} - \vec{k})^2+m_{\phi}^2}\frac{A_{i'j'}(k)}{\frac{\vec{k}^2}{2m_r^{i'j'}}-\mathcal{E}+2\delta m_{i'j'}}\, ,
\end{equation}
where the sum is over different $\chi_{i'}\chi_{j'}$ intermediate states and different interactions. Here $\mathcal{E}~=~{\vec{p}}\,^2/2m_r^{ab}$ is kinetic energy of incoming pair (at infinity), with $m_r^{ab}$ the reduced mass and $\vec{p}$ the CM three-momentum, $2\delta m_{ij}=m_{i}+m_{j}-(m_a+m_b)$ the mass splitting, $g_{ii'\phi}$ and $g_{j'j\phi}$ are the coupling constants and $N_{ij,i'j'}$ is the term containing normalization and combinatorial factors.
\par
To rewrite the expression above in the form of the Schr\"{o}dinger equations we make following redefinitions:
\begin{eqnarray}
\label{FT1}
A_{ij}(\vec{ p})&=&\left(\frac{\vec{ p}^2}{2m_r^{ij}}-\mathcal{E}+2\delta m_{ij}\right)\tilde\psi_{ij}(\vec{p}), \\
U_{ij}^0(\vec{r})&=&\int d^3 \vec{p}\, e^{i\vec{ p}\cdot\vec{r}} A_{ij}^0(\vec{ p},P_0), \\
\psi_{ij}(\vec{r})&=&\int d\vec{p}\, e^{i\vec{p}\cdot\vec{r}}\tilde\psi_{ij}(\vec{p}),
\end{eqnarray}
which allows us to rewrite Eq. (\ref{inteqn}) as a differential equation:
\begin{equation}
\label{schrr}
 -\frac{\partial^2}{2m_r^{ij}}\psi_{ij} (\vec{r})= U_{ij}^0(\vec{r})+\left(\mathcal{E}-2\delta m_{ij} \right)\psi_{ij} (\vec{r})+\sum_{i'j'\phi}V^\phi_{ij,i'j'}\psi_{i'j'}(\vec{r}),
\end{equation} 
where $\phi$ refers to the particle being exchanged (scalar, vector or axial vector boson). The potential has the form
\begin{equation}
\label{potentialr}
 V^\phi_{ij,i'j'}(r)= \frac{c_{ij,i'j'}(\phi)}{4\pi}\frac{e^{-m_\phi r}}{r}\, ,
\end{equation}
with $c_{ij,i'j'}(\phi)$ being coefficients depending on the couplings and states involved. An efficient way of computing them is explained in Appendix \ref{computationmethod} and the results in case of a system involving one spin $1/2$ Dirac fermion and/or two different Majorana spin $1/2$ fermions are summarized in Tab.~\ref{cTable}. Whether the potential is attractive or repulsive is hidden in the sign of the coefficients $c$ and depends on the interaction type. The exchange of scalars is always attractive in the spin singlet (i.e. overall plus sign), but can also be repulsive in the triplet. Vector and axial bosons can give attractive or repulsive forces, depending on the charges.
\TABLE{\label{cTable}
{\linespread{1.2}
{\small \begin{tabular}{l|ccc}
\toprule
 & \multicolumn{3}{c}{Spin singlet} \\
\midrule
$\phi:$ & scalar ($\Gamma=\dblone$) & vector ($\Gamma=\gamma_0$)  & axial ($\Gamma=\gamma_i \gamma_5$)\\ 
\midrule
$c_{+-,+-}$ & $g^2$ & $g^2$ & $-3g^2$  \\ 
$c_{++,++}$ & $g^2$ & $-g^2$ & $-3g^2$\\ 
$c_{ii,+-}$ & $\sqrt{2} |g_{i+}|^2$ & $\sqrt{2} |g_{i+}|^2$ & $-3\sqrt{2} |g_{i+}|^2$ \\ 
$c_{ij,+-}$ & $2\Rea(g_{i+} g_{j+}^*)$ & $2\Rea(g_{i+} g_{j+}^*)$ & $-6\Rea(g_{i+} g_{j+}^*)$\\ 
$c_{ii,jj}$ & $2|g_{ij}|^2+g_{ij}^2+g_{ij}^{*2}$ & $2|g_{ij}|^2-g_{ij}^2-g_{ij}^{*2}$ & $-3(2|g_{ij}|^2+g_{ij}^2+g_{ij}^{*2})$ \\ 
$c_{ij,ij}$ & $2|g_{ij}|^2+g_{ij}^2+g_{ij}^{*2}+4g_{ii}g_{jj}$ & $-2|g_{ij}|^2+g_{ij}^2+g_{ij}^{*2}$ & $-3(2|g_{ij}|^2+g_{ij}^2+g_{ij}^{*2})-12g_{ii}g_{jj}$ \\ 
$c_{+i,+i}$ & $|g_{i+}|^2+2g_{ii}g$ & $-|g_{i+}|^2$ & $-3|g_{i+}|^2-6g_{ii}g$ \\ 
$c_{+i,+j}$ & $g_{i+} g_{j+}^*+2g \Rea(g_{ij})$ & $-g_{i+} g_{j+}^*-2g i\Im(g_{ij})$ & $-3g_{i+} g_{j+}^*-6g \Rea(g_{ij})$ \\ 
$c_{ii,ii}$ & $4g_{ii}^2$ & 0 & $-12g_{ii}^2$ \\ 
$c_{ij,ii}$ & $4\sqrt{2}g_{ii}\Rea(g_{ij})$ & 0 & $-12\sqrt{2}g_{ii}\Rea(g_{ij})$ \\ 
\midrule
 &\multicolumn{3}{c}{Spin triplet} \\
\midrule
$c_{+-,+-}$  & $g^2$ & $g^2$ & $g^2$ \\ 
$c_{++,++}$  & $g^2$ & $-g^2$ & $g^2$\\ 
$c_{ii,+-}$  & 0 & 0 & 0\\ 
$c_{ij,+-}$  & $2i\Im(g_{i+}^* g_{j+})$ & $2i\Im(g_{i+}^* g_{j+})$ & $2i\Im(g_{i+}^* g_{j+})$\\ 
$c_{ii,jj}$ & 0 & 0 & 0\\ 
$c_{ij,ij}$  & $-(2|g_{ij}|^2+g_{ij}^2+g_{ij}^{*2})+4g_{ii}g_{jj}$ & $2|g_{ij}|^2-g_{ij}^2-g_{ij}^{*2}$ & $-(2|g_{ij}|^2+g_{ij}^2+g_{ij}^{*2})+4g_{ii}g_{jj}$\\ 
$c_{+i,+i}$  & $-|g_{i+}|^2+2g g_{ii}$ & $|g_{i+}|^2$ & $-|g_{i+}|^2+2g g_{ii}$\\ 
$c_{+i,+j}$ & $-g_{i+} g_{j+}^*+2g \Rea(g_{ij})$ & $g_{i+} g_{j+}^*-2g i\Im(g_{ij})$ & $-g_{i+} g_{j+}^*+2g \Rea(g_{ij})$\\ 
$c_{ii,ii}$  & 0 & 0 & 0\\ 
$c_{ij,ii}$  & 0 & 0 & 0\\
\midrule
Couplings:&\multicolumn{3}{c}{$g^\Gamma_{ij}\bar\chi_j\Gamma\chi_i\phi\quad (+h.c.\ {\rm iff}\ i\neq j),\qquad g^\Gamma_{i+}\bar\psi\Gamma\chi_i\phi\ +h.c.,\qquad 
g^\Gamma\bar\psi\Gamma\psi\phi, \qquad{\rm where}\quad \Gamma=\dblone,\gamma_0,\gamma_i\gamma_5$} \\
\bottomrule

\end{tabular} 
}}
\caption{List of all possible coefficients in the potential $V^\phi_{ij,i'j'}(r)$ for the Sommerfeld effect computation. The table includes any annihilation process 
involving one spin $1/2$ Dirac fermion (denoted by $+$ or $-$ depending on whether it is a particle or antiparticle) and/or two different Majorana spin $1/2$ fermions (denoted by $i$ and $j$), for an even 
partial wave. Couplings are defined in the last line for each $\Gamma$, where $\chi$ is a Majorana fermion, $\psi$ a Dirac field,
$\phi$ is the exchanged boson. When applied to the MSSM, where one needs to consider a chargino and one or two neutralinos, some of the coefficients 
vanish due to the CP conservation and couplings of type $g_{ii}$ are negligible. The overall ''$+$'' sign refers to an attractive force while ''$-$'' to a repulsive force. 
Note also that $c_{ij,kl}=c^*_{kl,ij}$.}
}
\par
Notably in the approach outlined here we can split different interaction types (i.e. mediated by different bosons or with different couplings etc.) within a ladder diagram and consider them separately. The trade-off is that every possible intermediate two-particle state will lead to one equation, so in such a general case we will need to consider a set of coupled Schr\"{o}dinger equations. These equations are inhomogeneous; however, since the Sommerfeld enhancement factorizes out from the annihilation matrix, as it does not depend on the final states but enters only as a distortion of the incoming wave function, one can compute it by solving the associated homogeneous, using the partial waves decomposition, as described in \cite{Iengo:2009ni}. We will be interested only in the s-wave, but it is straightforward (though not always easy numerically) to extend the analysis to higher partial waves.
\par
To reduce these equations in a form more suitable for numerical calculations, after the partial wave decomposition we define the reduced radial wave function $\varphi(x)$ as 
\begin{equation}
 R^{ij}_{p,l}(r)=Np\frac{\varphi^{ij}_l(x)}{x}\, , \qquad\quad x=pr\, ,
\end{equation} 
where $N$ is some normalization constant and $p$ is the value of the CM three-momentum for one of the incoming particles ($a$ or $b$). Since we restrict to the s-wave case $l=0$, we will drop the $l$ index from now on. From (\ref{schrr}) one get then set of equations\footnote{This form is most suitable for numerical solutions, while often it is presented not in terms of the CM momentum, but rather relative velocity of incoming particles, being equal to $v=p/m^{ab}_r$.} for the $\varphi^{ij}(x)$:
\begin{equation}
\label{schrx}
 \frac{d^2\varphi^{ij}(x)}{dx^2}+\frac{m_r^{ij}}{m_r^{ab}}\left[\left(1-\frac{2\delta m_{ij}}{\mathcal{E}}\right)\varphi^{ij}(x)+\frac{1}{\mathcal{E}}\sum_{i'j'\phi}V^\phi_{ij,i'j'}(x)\varphi^{i'j'}(x)\right]=0\, .
\end{equation}
To obtain the enhancement we need to solve this set of equations with appropriate boundary conditions. In $x=0$ they are set by the requirement that the solution is regular. In $x\rightarrow\infty$ the solution has to describe one incoming $\chi_a\chi_b$ state and all the possible $\chi_i\chi_j$ states that can be produced in the ladder. For the latter there can be two cases:
\begin{enumerate}
 \item $2\delta m_{ij} < \mathcal{E}$ - there is enough energy to produce on-shell states $\chi_i\chi_j$,
\item $2\delta m_{ij} > \mathcal{E}$ - there is \textit{not} enough energy; states $\chi_i\chi_j$ are off-shell.
\end{enumerate}
The radial wave functions behave at infinity as: 
\begin{equation}
 R^{ab}(r)\rightarrow \frac{C_1^{ab}}{2i}\frac{e^{ik_{ab} r}}{r}-\frac{1}{2i}\frac{e^{-ik_{ab} r}}{r}\, ,
\end{equation}
for the incoming pair and
\begin{equation}
R^{ij}(r)\rightarrow \Biggl\lbrace 
 \begin{array}{ll}
 \frac{C_2^{ij}}{2i}\frac{e^{ik_{ij} r}}{r} & \rm{if\ on-shell}, \\
 \frac{D_2^{ij}}{2i}\frac{e^{-|k_{ij}| r}}{r} & \rm{if\ off-shell},
\end{array} 
\end{equation}
for every other intermediate state $\chi_i\chi_j$. We use the normalization of the wave function to be such, that for the non-interacting case $R_{ab}=\sin (k_{ab} r)/r$ (for details see \cite{Iengo:2009ni}). After changing the variable to $x=k_{ab}\cdot r$ and defining $q_{ij}={m_r^{ij}}/m_r^{ab}\cdot\left(1-2\delta m_{ij}/\mathcal{E}\right)$ 
we get set of boundary conditions for the reduced wave functions at $x\rightarrow \infty$:
\begin{equation}
i\varphi^{ab}-\partial_x\varphi^{ab}=-e^{-ix},
\end{equation} 
\begin{equation}
\biggl\lbrace 
 \begin{array}{ll}
 i\sqrt{q_{ij}} \varphi^{ij}-\partial_x\varphi^{ij}=0 & \rm{if\ on-shell}, \\
 \sqrt{-q_{ij}} \varphi^{ij}+\partial_x\varphi^{ij}=0 & \rm{if\ off-shell}.
\end{array} 
\end{equation}

To check our numerics we can use the unitarity condition, saying that:
\begin{equation}
1=|C_1^{ab}|^2+\sum_{ij}\sqrt{q_{ij}}|C_2^{ij}|^2.
\end{equation}
Since the set of boundary conditions depends on the initial particles masses and energy, the solutions to the Schr\"{o}dinger equations, i.e. the wave functions, depend on the initial conditions of the incoming pair $\chi_a\chi_b$: to be precise we should call them then $\varphi_{(ab)}^{ij}(x)$. After solving (\ref{schrx}) the (co-)annihilation cross section of the pair $\chi_a\chi_b$ is determined, up to the kinematical factor, by
\begin{equation}
 \sigma_{(ab)}\propto\sum_{ij} S^{ij}_{(ab)}\cdot|A^0_{ij}|^2,
\end{equation} 
where the enhancement factors with our normalization are
\begin{equation}
 S^{ij}_{(ab)}=|\partial_x\varphi^{ij}_{(ab)}|^2_{x=0}\, .
\end{equation} 
\par
It is important to note, that the computation of the enhancement depends on the spin state of the initial 2-body state (see Tab. \ref{cTable}). This means that one needs to project each annihilation cross section into two parts, one for the singlet and one for the triplet initial spin state, multiply each of them by a different enhancement factor. The method we used to do so is described in Appendix \ref{projectors}.
\par
Would one need to include higher partial waves, one has to compute $c_{ij,i'j'}(\phi)$ coefficients for odd partial waves, add a centrifugal term to the equations (\ref{schrr}) and (\ref{schrx}) and modify the expression for the enhancement, see \cite{Iengo:2009ni}.

\section{Relic density computation}
\label{relic}

The multi-state scenario we have just introduced for the Sommerfeld enhancement is a typical setup in which the computation of
the thermal relic density for the lightest of such states is strongly affected by co-annihilations~\cite{Binetruy:1983jf,Griest:1990kh}.
To model this effect, we will follow here the approach of Ref.~\cite{Edsjo:1997bg}, appropriately modified to include 
the Sommerfeld enhancements; such treatment will not involve approximations in the way thermally averaged
annihilation cross sections are computed, while the density evolution equation is solved fully numerically.  We
summarize briefly here the steps involved in the computation. 

Let $\chi_1$, $\chi_2$, ... $\chi_M$,  be a set of $M$ particles, each with mass $m_i$ (the ordering is such that $m_1 \le m_2 \le \dots \le m_M$)
and number of internal degrees of freedom $h_i$, sharing a conserved quantum number so that: {\sl i)} if kinematically
allowed, inelastic scatterings on SM thermal bath particles can turn each of these states into another,  and {\sl ii)}  $\chi_1$ is stable.
If the mass splitting between the heaviest and the lightest is comparable with $m_1/20$, roughly speaking
the freeze out temperature for a WIMP, all these states have comparable number densities at decoupling
and actively participate to the thermal freeze out. A system of $M$ coupled Boltzmann equations can be written 
to trace the number density $n_i$ of each single state; since, however, after freeze out all heavier states
decay into the lightest, one usually solves a single equation written for the sum of the number densities, 
$n = \sum_i n_i$, i.e.~\cite{Edsjo:1997bg}:
\begin{equation}
\label{eq:1eq}
\frac{dn}{dt} + 3\,H\,n = - \langle\sigma_{\rm{eff}} v\rangle
\left[n^2 - (n^{eq})^2 \right]\, ,
\end{equation}
with the effective thermally averaged annihilation cross section: 
\begin{equation}
\label{eq:sum}
\langle\sigma_{\rm{eff}} v\rangle = \sum_{i,j} \langle\sigma_{ij} v_{ij}\rangle
\frac{n_i^{eq}}{n^{eq}} \frac{n_j^{eq}}{n^{eq}}\, 
\end{equation}
written as a weighted sum over the thermally averaged annihilation cross section for processes of the type
$\chi_i + \chi_j \rightarrow X$ (in the dilute limit, two-body initial state processes dominate): 
\begin{equation} 
\label{eq;sigmav}
\langle\sigma_{ij} v_{ij}\rangle =
\frac{1}{n_i^{eq}\, n_j^{eq}}
\sum_{X}
\int \frac{d^3p_i}{2\,E_i}  \frac{d^3p_j}{2\,E_j}  \frac{d^3p_X}{2\,E_X}  \delta^4(p_i+p_j-p_X) f_i^{eq}(p_i) f_j^{eq}(p_j)\, |A_{ij\rightarrow X}|^2 \, ,
\end{equation}
where the sum here is on the set $X$ of allowed SM final states (normally only two-body final states are considered) and $d^3p_X/2\,E_X$ stands 
symbolically for the integration over the phase space in the final state. In the equations above $n_i^{eq} = \int d^3p f_i^{eq}(p_i)$ is the thermal equilibrium 
number density for the species $i$, while $n^{eq}$ the sum of  $n_i^{eq}$ over all states. There are two main assumptions which allow to rewrite the
system of $M$ coupled  Boltzmann equations as the single equation for $n$, in analogy to the case one writes for one single WIMP, with the usual 
term of dilution by volume on the l.h.s. and the depletion and replenish terms on the r.h.s. The factorization of the individual terms in the sum 
of Eq.~(\ref{eq:sum}) is possible if one assumes that the shape of phase space densities for each particle $\chi_i$ follows the shape of the corresponding thermal 
equilibrium phase space density, namely  $f_i(p_i,t)= c(t) \cdot f_i^{eq}(p_i,t)$ (with the coefficient $c$ depending on time but not on momentum); this is the case
if the so-called kinetic equilibrium is maintained, i.e. if scattering processes of the kind $\chi_i + X_l \rightarrow \chi_k + X_m$ (with $k$ equal to $i$ or different 
from it) on SM thermal bath states $X_l$ and $X_m$ have a rate which is larger than the Universe expansion rate $H$. The amplitudes for scattering 
and annihilation processes are usually comparable since the two can be related via crossing symmetry; on the other hand scatterings are stimulated by
thermal bath states, which are relativistic and hence whose number density is much larger than the number density of the particle $\chi$ at the freeze-out
temperature $\sim m_1/20$. Hence kinetic decoupling usually takes place much later than chemical freeze-out (chemistry here indicates number density
changing processes). Since, as we will show below, within the particle physics setup we are considering the Sommerfeld factors are of order $O(1)$, this is typically the case even when the depletion term takes
the Sommerfeld effect into account. In case of a resonance the kinetic decoupling can indeed happen when the
depletion term is still active, as we will discuss in Section \ref{kineticdecoupling}. The kinetic decoupling can have a much larger impact in more extreme scenarios, as discussed in Ref. \cite{Dent:2009bv}.
\par
The second assumption which is implicit in Eq.~(\ref{eq:1eq}) is that one takes $n_i/n \simeq n_i^{eq}/n^{eq}$, a quantity which in the Maxwell-Boltzmann approximation
for equilibrium phase space densities, as appropriate for WIMPs, i.e. $f_i^{eq}(p_i,t) = h_i/(2\pi)^3 \exp (-E_i/T)$, is proportional to the number of internal degrees of freedom $h_i$ and is exponentially suppressed with the mass splittings. Analogously to the first assumption, this approximation is valid in case inelastic scatterings are active for the whole phase in which the depletion term is relevant.

Usually the thermally averaged annihilation cross sections in Eq.~(\ref{eq;sigmav}) are computed at the lowest order in perturbation theory taking tree-level amplitudes.
Here we will include the Sommerfeld effect introducing the rescaling $|A_{ab}|^2=\sum_{ij}S^{ij}_{(ab)}\,|A_{ij}^0|^2$, with $S^{ij}_{(ab)}$ as computed in previous Section. Actually, since the 
effect can be interpreted as a rescaling in the wave function of the incoming pair, $S^{ij}_{(ab)}$ does not depend on the final state $X$ and can be factorizes out of the total 
annihilation rate $W_{ij}$. Following the same steps of Ref.~\cite{Edsjo:1997bg} and adopting an analogous notation, one finds:
\begin{equation} 
\label{eq:sigmav2}
  \langle \sigma_{\rm{eff}}v \rangle = \frac{\int_0^\infty dp_{\rm{eff}} p_{\rm{eff}}^2 SW_{\rm{eff}}(p_{\rm{eff}},T) K_1 \left(
  \frac{\sqrt{s}}{T} \right)} {m_1^4 T \left[ \sum_i \frac{h_i}{h_1}
  \frac{m_i^2}{m_1^2} K_2 \left(\frac{m_i}{T}\right) \right]^2}\,,
\end{equation}
where we have defined:
\begin{equation} 
\label{eq:sweff}
  SW_{\rm{eff}}(p_{\rm{eff}},T) = \sum_{ab}\sum_{ij}\frac{p_{ij}}{p_{\rm{eff}}}
  \frac{h_i h_j}{h_1^2} S_{(ab)}^{ij}(p_{\rm{eff}},T) W_{ij}(p_{\rm{eff}})
\end{equation} 
with:
\begin{equation} 
   p_{ij} = \frac{1}{2}  \sqrt{\frac{[s-(m_{i}-m_{j})^2][s-(m_{i}+m_{j})^2]}{s}}
\end{equation} 
and $p_{\rm{eff}} = p_{11} = 1/2 \sqrt{s-4m_{1}^2}$. The explicit dependence of $S^{ij}_{(ab)}$ on $T$ stems from the fact that there may be an explicit dependence
of the mass of the long-range force carriers on temperature, see the discussion below; its dependence on the relative velocity of the annihilating pairs
has been rewritten instead in term of the integration variable, i.e. the effective momentum $p_{\rm{eff}}$.
 
\section{Implementation of the method to the MSSM}

We apply the general method outlined in the previous Sections to one of the best motivated and most extensively studied scenarios, namely the case of neutralino dark 
matter in the MSSM. Since we will focus the discuss to cases in which the sfermion sector is not playing any relevant role, to simplify the discussion and underline better
which are the key parameters, we choose actually to refer to the SUSY framework usually dubbed "Split Supersymmetry"~\cite{ArkaniHamed:2004fb,Giudice:2004tc}.
This indicates a generic realization of the SUSY extension to the SM where fermionic superpartners feature a low mass spectrum 
(say at the TeV scale or lower), while scalar superpartners are heavy, with a mass scale which can in principle range from hundreds of TeV up to the GUT or the 
Planck scale \cite{ArkaniHamed:2004fb}, a feature which can occur in wide class of theories, see, e.g.~\cite{ArkaniHamed:2004yi,Antoniadis:2004dt,Kors:2004hz}.
We will also leave out of our discussion the Gluino and the gravitino, supposing they are (moderately) heavy, focussing the analysis on neutralinos and 
charginos.\footnote{The Sommerfeld effect does play a role in the 
pair annihilation rate of charged sfermions or gluinos, and so it does affect the computation of the neutralino relic density in case of sfermion or gluino co-annihilations; there
is however no long-range interaction turning a pair of these particle into a neutralino pair, and hence, contrary to case discussed in this paper, the different annihilation can 
be treated separately and the calculation is simplified. An analysis dedicated to these co-annihilation channels will be presented in Ref.~\cite{Hryczuk:2011tq}.}
The MSSM contains four neutralinos, spin~1/2 Majorana fermions arising as the mass eigenstates from the superposition of the two neutral Gauginos, 
the Bino $\tilde B$ and the Wino $\tilde W^3$,  and the two neutral Higgsino fields $\tilde H^0_1$ and $\tilde H^0_2$:
\begin{equation}
  \tilde{\chi}^0_i = 
  N_{i1} \tilde{B} + N_{i2} \tilde{W}^3 + 
  N_{i3} \tilde{H}^0_1 + N_{i4} \tilde{H}^0_2\, ;
\end{equation}
in the scheme we are considering the lightest neutralino is automatically the lightest SUSY particle (LSP) and, possibly, a good WIMP dark matter candidate. 
The coefficients $N_{ij}$, obtained by diagonalizing the neutralino mass matrix, are mainly a function of the Bino and the Wino mass parameters $M_1$ and $M_2$, 
and of  the Higgs superfield parameter $\mu$, while depend rather weakly on $\tan\beta$, the ratio of the vacuum expectation values of the two neutral components 
of the SU(2) Higgs doublets, which appears in the off-diagonal terms of neutralino mass matrix. The two MSSM charginos are instead spin 1/2 Dirac fermions 
obtained as mass eigenstates from one charged Wino state and one charged Higgsino; the chargino composition is again mainly set by the relative weight of
$M_2$ and $\mu$. 

The long-range forces driving a sizable Sommerfeld effect are the weak force on neutralinos and charginos  due to the exchange of $W^{\pm}$ and $Z^0$, and the
electromagnetic force with photon exchanges; we also include the effect of the charged Higgs  $H^{\pm}$ and the two CP-even neutral Higgs $H^0_{1}$ and
$H^0_{2}$, which however play a minor role. The last MSSM scalar, the CP-odd neutral Higgs $H^0_{3}$, does not give rise to any contribution for s-wave annihilations 
in the non-relativistic regime. Since the $\tilde B$ is not charged under $SU(2)_L$, a large Bino component in the lightest neutralino drastically reduces the relevance 
of the Sommerfeld effect; we will then consider only the case of  $M_1\gg M_2$. Higgsinos and Winos have a pair annihilation cross section into $W$ and $Z$ bosons
which is fairly large, much larger than the standard reference value for thermal relics of $3\cdot 10^{-26}\;{\rm cm}^3\,{\rm s}^{-1}$ if their mass is at around 100~GeV.
Going however for more massive neutralinos, namely around 1.1~TeV for Higgsinos and 2.2~TeV for Winos, the standard tree-level 
calculation of the thermal relic density gives a result which is compatible with the measured value for the energy density in CDM. This heavy mass regime is also the
one in which the Sommerfeld enhancement condition of mass of the particle much heavier than mass of the force carrier is realized for weak interactions; hence relevant
corrections to the tree-level estimate of the relic abundance may arise~\cite{Hisano:2006nn,Cirelli:2008jk}. All the pair annihilation processes we will need 
to consider are dominated by their s-wave contribution (unlike the case of the pair annihilation of Binos into a fermion-antifermion); moreover the enhancement for 
the higher partial waves is smaller, due to repulsive centrifugal term in the Schr\"{o}dinger equation.\footnote{This statement is not true only in the case of resonances. In the 
Coulomb case, when analytical expressions can be derived, every partial wave is enhanced by the factor $\sim (\alpha/v)^{2l+1}$, so for higher $l$ the resonant enhancement 
is higher and narrower. However, since one has to integrate over the thermal distributions, the resonances are smeared and net impact of the higher partial waves is very 
small.}  We can then safely assume that only the s-wave Sommerfeld effect is relevant.

As we stressed in Section~\ref{sec:somn}, the value of the mass splitting among the different states is crucial in the analysis. When calculating the mass spectrum we include
the radiative corrections to the neutralino and chargino masses due to gauge boson loops~\cite{Pierce:1994ew,Cheng:1998hc} (tree-level neutralino and chargino masses
are degenerate in the pure Higgsino or pure Wino limits) and the thermal dependence of the Higgs VEVs. 
Thermal effects are also important when considering vector boson masses, as already stressed in the analysis of Ref.~\cite{Cirelli:2007xd}. There are two thermal 
effects which we need to include: First, we approximate the scaling of the VEV $v$ with temperature as~\cite{Dine:1992wr}:
\begin{equation}
 v(T)=v\Rea\left(1-\frac{T^2}{T_c^2}\right)^{1/2}\, ,
\end{equation}
with the critical temperature $T_c$ depending on the Higgs mass (as in Ref.~\cite{Cirelli:2007xd}, we will assume $T_c=200$~GeV). Second, we consider the contribution to 
gauge boson masses, or, in more appropriate terms, to their propagator pole, due to  the screening by the thermal plasma; this effect can be approximated by adding the 
so-called Debye mass~\cite{Gross:1980br,Weldon:1982bn}, which in the SM and  at $T\gg m_{W,Z}$ are:
\begin{equation*}
 \Delta m^2_\gamma=\frac{11}{6}g_Y T^2\, , \qquad \Delta m^2_{W,Z}=\frac{11}{6}g_2 T^2\, .
\end{equation*} 
We will assume that these expressions are valid also for the MSSM and at lower temperatures; the first assumption is justified by the fact that the contributions from the
additional states are small since these particles are heavy, the latter has a negligible impact on the relic density calculation. The two effects  introduce a correction to the Sommerfeld factors which is non-negligible, but still quite small, since they are important mostly for high temperatures, at which the Sommerfeld effect is anyway negligible. 

Having computed with high accuracy the mass spectra, we implement the procedure for the relic density, considering a system of coupled equations which includes
all states with small mass splitting compared to the LSP. In practice this reduced to including two Majorana and one Dirac fermion when $\mu < M_2$ (i.e. two neutralinos and one chargino, mainly Higgsino-like) and one Majorana and one Dirac fermion when $\mu > M_2$  (i.e. one neutralino and one chargino, mainly Wino-like).  Additionally, since the Sommerfeld enhancements depend on the total spin of the initial pair, we considered each case separately if needed. Again, in practice this is important only if the incoming particles are one neutralino and one chargino: two identical Majoranas cannot form a s-wave spin triplet state, the triplet of two different Majoranas has suppressed annihilations (it cannot annihilate to $W^+W^-$), and for two charginos the effect comes dominantly via $\gamma$ exchange, which is vector and both singlet and triplet computations coincide.

\section{Results and discussion}

Having introduced the method and particle physics framework, we present results for the Sommerfeld factors and the impact of the effect on the relic density of the neutralino. 
As already explained, $M_2$ and $\mu$ are the critical parameter in this problem; we will consider an approach with parameters set at the low energy scale (rather than at a 
GUT scale as often done) and let them vary freely. The other MSSM parameters are kept fixed: besides the sfermion sector which is assumed to be heavy, and the Bino which
is also decoupled with the artifact of setting $M_1=100 M_2$, we assume $\tan \beta=30$ and an Higgs sector with a light SM-like Higgs and all other states that are heavy,
as expected in split SUSY. The value of $\tan\beta$ has a very modest impact on results and considering other values does not bring other information. In the same way, 
the other Higgs sector parameters  have very little relevance. For any given point in this two-dimensional parameter space, the relic density calculation is performed computing 
first the coefficients $S_{ij}(v,T)$ numerically, adopting a standard adaptive $5^{\rm{th}}$ order Runge-Kutta method with shooting for solving the boundary value problem \cite{Press}.

\subsection{Sommerfeld factors and the effective cross section}

For any given point in the MSSM parameter space leading to a lightest neutralino $\chi_1^0$ which is heavy and has a large Wino or Higgsino fraction there are several Sommerfeld enhancement factors $S^{ij}_{(ab)}(v,T)$ needed for a relic density computation. The interplay between the different contributions to the thermal averaged effective cross section is in most cases non-trivial. In Fig. \ref{plot:enh} we give an example of $S^{ij}_{(ab)}(v,T)$ for the case when the neutralino is nearly purely $\tilde W^0$, just to give a intuition of what magnitude of enhancement factors we deal with. In this case in set of coupled equations (\ref{schrx}) we have $N\le 2$, i.e. only neutralino and chargino coupled together.
\FIGURE{\label{plot:enh}
		\includegraphics[scale=0.75]{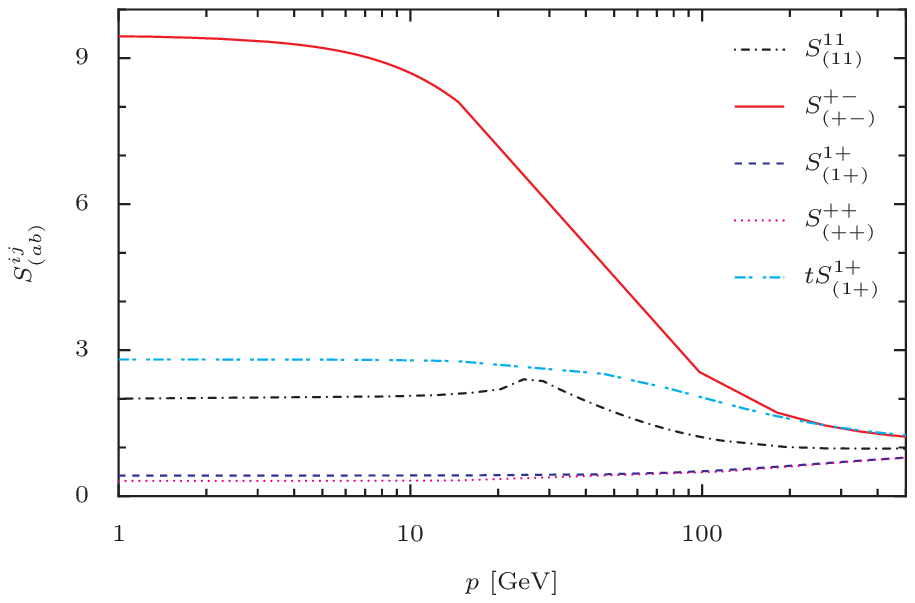}
		\includegraphics[scale=0.75]{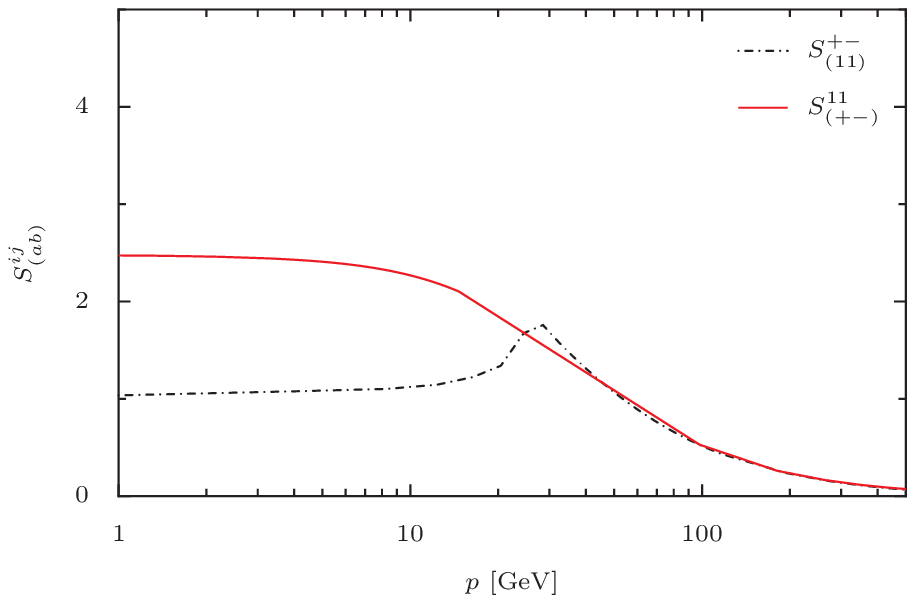}
		\caption{Example enhancement factors for $M_2=2.53$ TeV, $\mu=5$ TeV and $x=20$ (corresponding to $T\approx~126$~GeV). Subscripts refer to the incoming state with 1 - $\chi^0_1$ and $\pm$ - $\chi^\pm$, while superscripts to the annihilating one. The left panel shows Sommerfeld factors when the annihilating pair is the same as the incoming one, while right panel when they are different. In the latter case these factors approach 0 because at high momenta there is no Sommerfeld enhancement and the annihilation amplitude is suppressed because it can only be obtained at 1-loop level at least. All factors are computed for singlet spin state annihilation, except one indicated as $tS$ being for triplet.}}
One can see that the enhancements are of order $O(1)$ and very quickly go to 1 or to 0 in higher momenta and that most channels are attractive while two are repulsive, namely $\chi^+\chi_1^0$ in the singlet spin state and $\chi^+\chi^+$. Also, as expected, one gets the highest enhancement factor for the $\chi^+\chi^-$ channel, since there is a long range nearly\footnote{Due to the thermal corrections photon acquires small mass, which makes the potential to be Yukawa type, and the enhancement factor saturates at small $p$.} Coulomb interaction present. For the cases with two coupled equations and lighter states incoming we see a resonance in the value of momentum corresponding to the mass splitting. This is easy to understand, since for this energy the heavier states are produced nearly at rest, so they feel large effect coming from for instance $\gamma$ exchange. This is perfectly consistent with what was found in similar cases using analytical approximations \cite{Slatyer:2009vg}.
\par
We are now ready to compute $\sigma_{\rm{eff}}$ and solve the Boltzmann equation as discussed in Section~\ref{relic}. Values for the effective annihilation cross section including the Sommerfeld effect and without it are shown in Fig. \ref{plot:sigmaeff} for sample cases of Wino- and Higgsino-like neutralino.

\FIGURE{\label{plot:sigmaeff}
 		\includegraphics[scale=0.70]{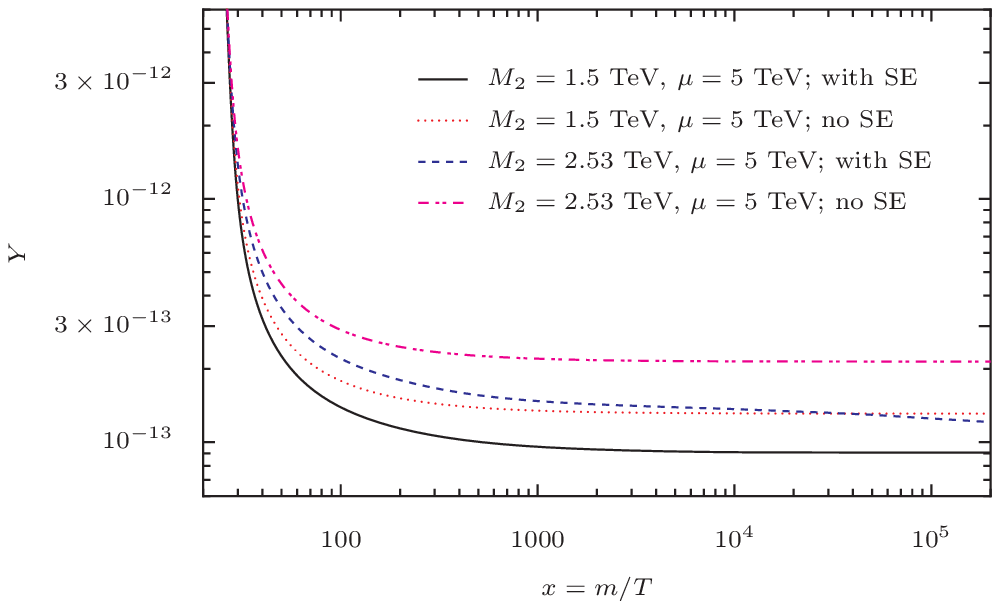}
 		\includegraphics[scale=0.70]{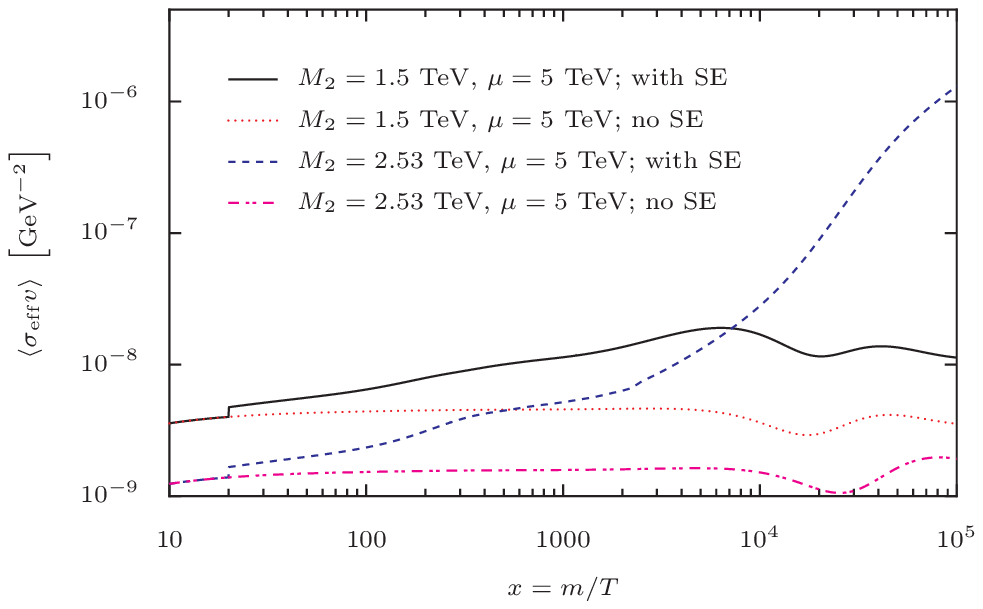}
 		\includegraphics[scale=0.69]{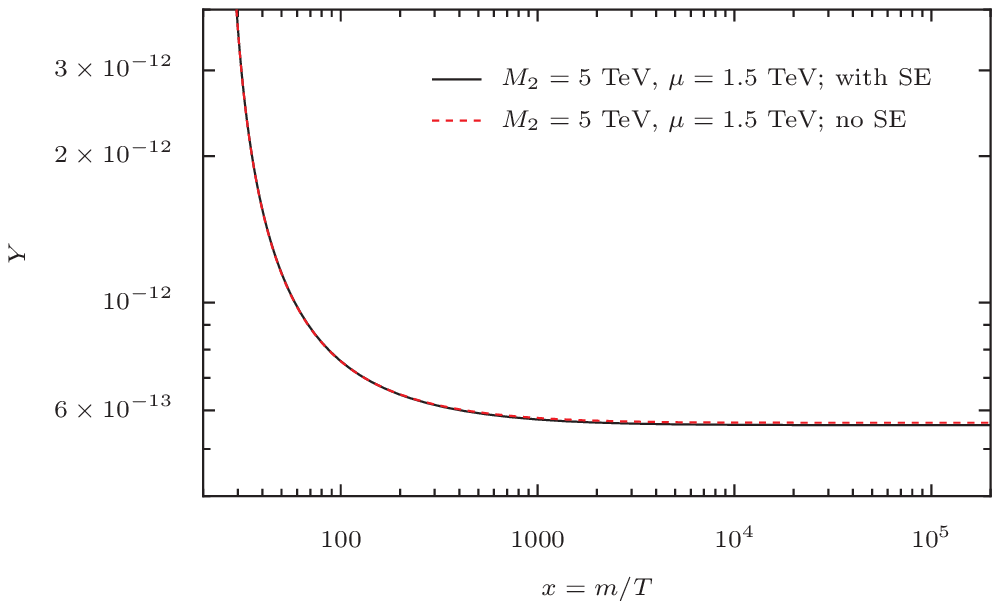}
 		\includegraphics[scale=0.69]{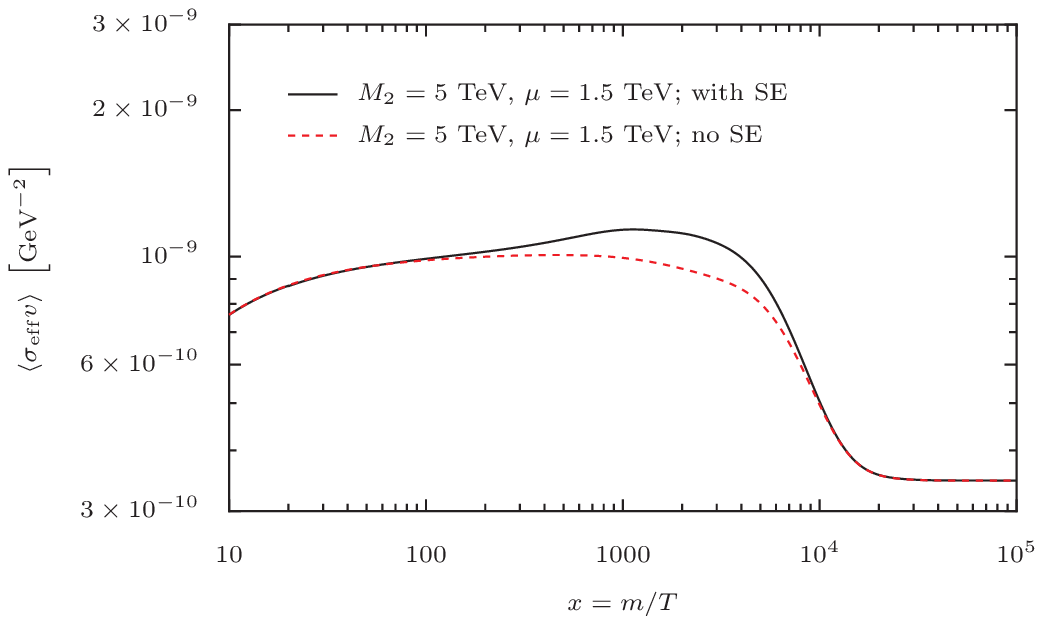}
		\caption{Number densities in units of entropy density $Y$ and effective cross sections for the Wino-like \textit{(top)} and Higgsino-like neutralino \textit{(bottom)}. In the Wino case two set of parameters are presented - generic one and close to the resonance.}}
\par
In the Higgsino-like case the effect is very mild. The Sommerfeld enhancement of the effective cross section becomes relevant only in the small velocities regime, when the depletion term in the Boltzmann equation is marginally effective. This gives rise to a change in the relic density which is at most at the level of a few per cent. In the Wino-like case the picture looks much more interesting. The net effect on the both yield and $\sigma_{\rm{eff}}$ is clearly visible and can become even very large in the parameter range where large resonance effects occur.
\par
The reason for the difference in the behaviour of the Sommerfeld effect in the Wino- and Higgsino-like case comes mainly from the mass splitting between the lightest neutralino and chargino, which is typically much smaller in the Wino case. The Sommerfeld effect for the neutralinos relies mostly on a production of nearly on-shell charginos in the loop. Also the efficiency of co-annihilation, and subsequently the effect of the Sommerfeld enhancement coming from its impact on co-annihilating particles, strongly depends on the mass splitting. Hence, the larger the mass splitting the smaller the overall impact on the thermally averaged cross section.\footnote{Note however, that in general Sommerfeld factors themselves are not monotonic functions of the mass splitting, see e.g. Ref. \cite{Slatyer:2009vg}.}

\subsection{Relic density}

In Fig.~\ref{plot:oh2scan} and~\ref{plot:enhscan} we show results for the neutralino relic density in the plane $M_2$ versus $\mu$, varying these parameters in the range
500~GeV to 5~TeV, a region in which the thermal relic density varies from values much below the cosmologically preferred one to much above it. In the left panel  of 
Fig. ~\ref{plot:oh2scan}, we are plotting results assuming the usual tree-level approximation for the pair annihilation amplitude, while in the right panel those including the full 
treatment of the Sommerfeld effect are shown. Most manifestly, there is a sharp shift in the Wino-like region consistent with the 7-year WMAP data to heavier masses;
when including the Sommerfeld enhancement a pure Wino is found to have $\Omega h^2=0.11$ for a mass of about 2.8~TeV.  Much milder changes place in the Higgsino-like region; the relic density is practically unchanged when 
considering models in the cosmologically interesting band. The relic density decrease is even larger in the ultra-heavy regime, with the Sommerfeld effect becoming larger and larger as 
the gauge boson masses are becoming less important; this regime would however be consistent with cosmology only invoking some extra ingredient, such as, e.g., a dilution 
effect via late 
entropy production or the decay of the neutralino into a lighter state, such as a gravitino or an axino, which is the true LSP (although viable, both scenario require large 
fine-tuning).  The results found here for pure Winos and pure Higgsino are analogous to those of earlier works in Ref.~\cite{Hisano:2006nn} and Ref.~\cite{Cirelli:2007xd} in the 
same limiting cases; there are small quantitative differences stemming from the fact that we have identified more annihilation channels that need to be treated separately,
with the Sommerfeld factor depending on the initial spin state of the annihilating particle pair,  we have found a different coefficient in the axial vector exchange, i.e. that the axial vector has an additional $-3$ factor with respect to the vector (in agreement 
with the result in Ref.~\cite{Drees:2009gt}, and we have probably a better control of the numerical solution of the Boltzmann Eq. having implemented our full treatment in 
the \ds\ numerical package (the slight difference in numerical results between \cite{Hisano:2006nn} and \cite{Cirelli:2007xd} are instead probably mainly due to the thermal corrections implemented here following \cite{Cirelli:2007xd}).

\FIGURE{\label{plot:oh2scan}
 		\includegraphics[scale=0.55]{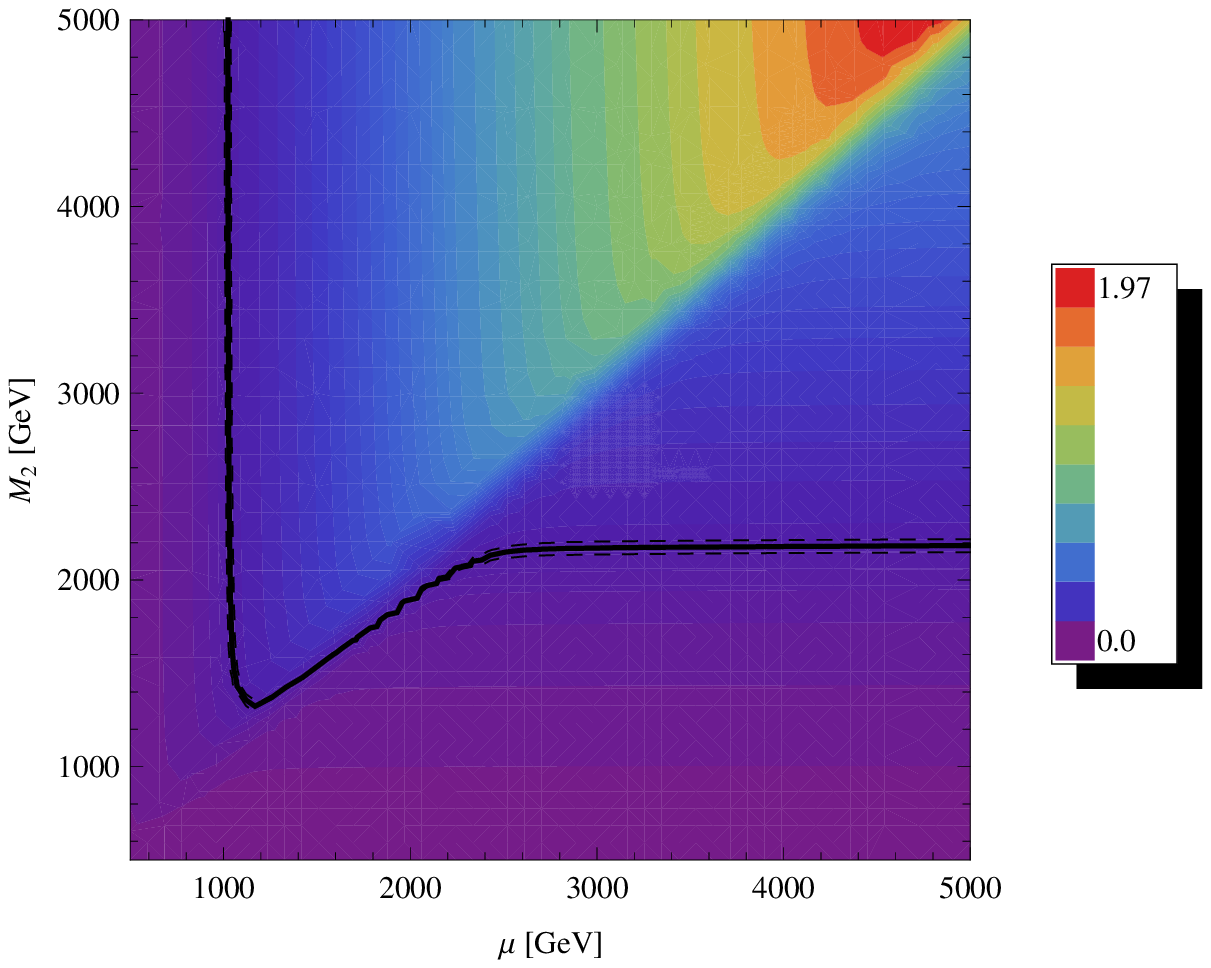}
 		\includegraphics[scale=0.55]{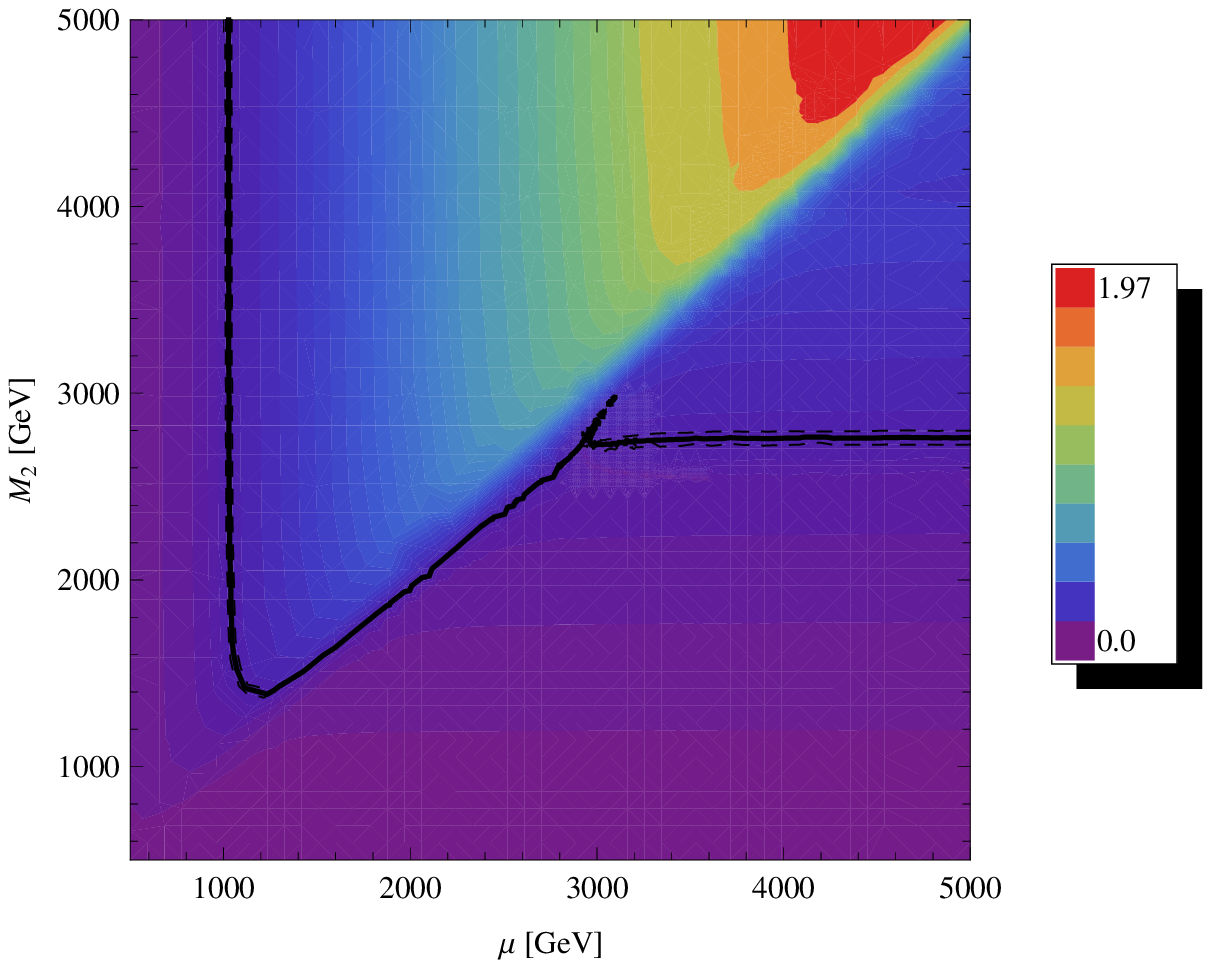}
		\caption{Relic density $\Omega h^2$ in the $\mu$-$M_2$ plane for perturbative case \textit{(left panel)} and with Sommerfeld effect included \textit{(right panel)}. The brighter the colour the higher $\Omega h^2$ and the colour scale is linear. The solid line and dashed lines correspond to the central value and the 1~$\sigma$ error bar for relic density consistent with the 7-year WMAP data, $\Omega h^2 = 0.1123\pm0.0035$.} }

\FIGURE{\label{plot:enhscan}
 		\includegraphics[scale=0.55]{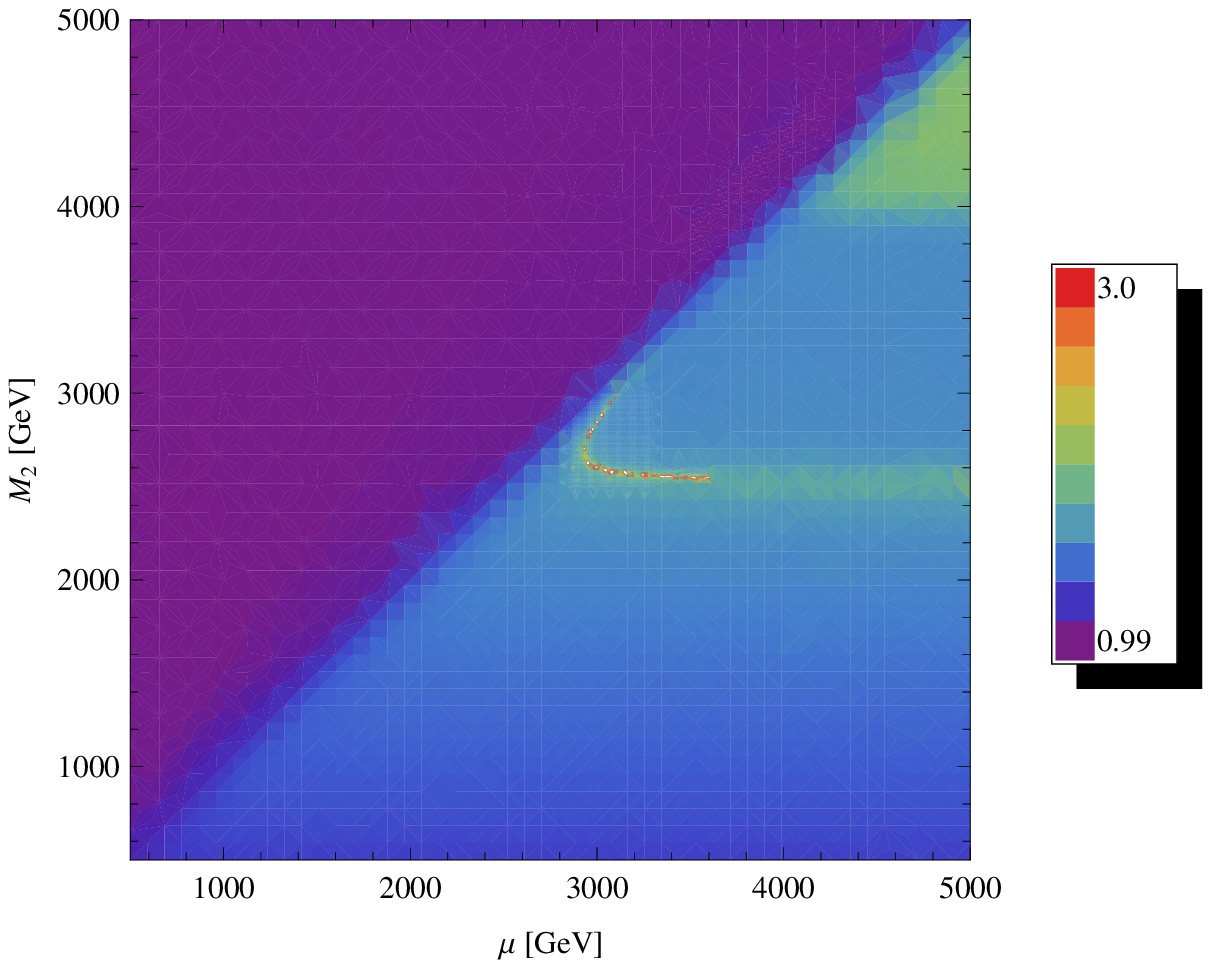}
 		\includegraphics[scale=0.55]{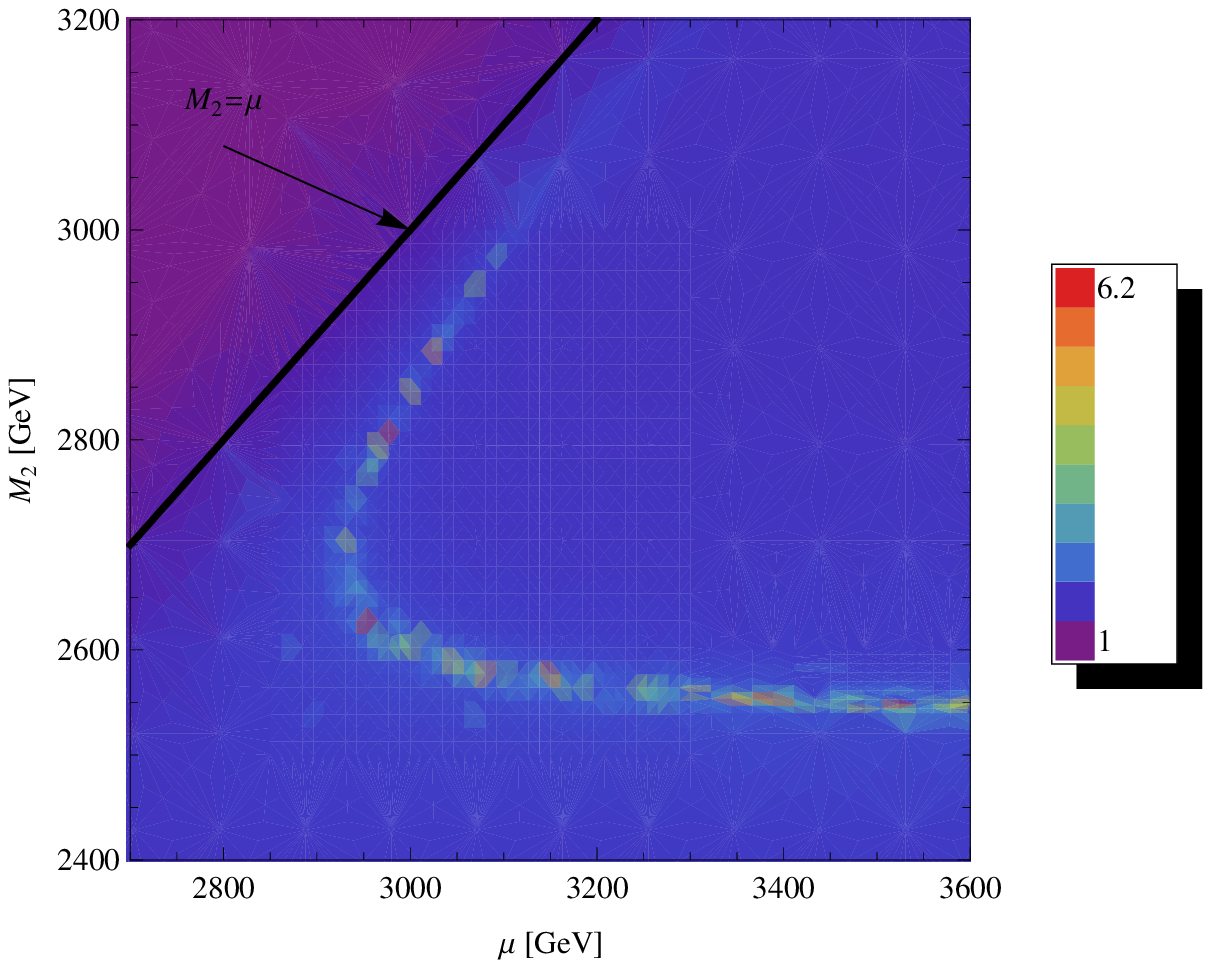}
		\caption{\textit{Left panel}: ratio of relic densities without and with Sommerfeld effect, $(\Omega h^2)_0/(\Omega h^2)_{\rm{SE}}$. \textit{Right panel}: the same ratio but focussing on the inner part of the resonant region. The colour scale is linear.} }

Another region showing interesting results is the band connecting the pure Higgsino to the pure Wino limit, towards $M_2 \sim \mu$ but still with a predominant
Wino component. Features in this region are more clearly seen in Fig.~\ref{plot:enhscan} where we show the ratio between the relic density computed with tree-level amplitudes
to the one with the full non-perturbative treatment. A thin ``resonance'' slice appears in the plane, starting for pure Winos with mass  $m_{\chi^0}\approx 2.5$~TeV and extending  
to heavier masses into the region with a sizable Higgsino fraction, where the thermal relic density becomes consistent with observations. 
The value of the mass we find for a pure Wino is precisely the one saturating  Eq. (\ref{longrange}), i.e.:
\begin{equation}
 \frac{1}{m_W}\approx \frac{1}{\alpha m_{\chi^0}}\, ,
\end{equation} 
with $\alpha$ as computed in the vertex for pure Winos $\tilde{W}^0\,\tilde{W}^+\, W^-$.  This means that the observed resonance is due to the possibility of creating the loosely bound state, occurring when the Bohr radius coincides with the interaction range. 
It also explains why, when we increase the Higgsino fraction, the resonance deviates to higher masses: a larger Higgsino fraction implies an increase in the mass splitting between 
lightest neutralino and chargino (since one goes from a mass splitting dominated by the radiative corrections, about 170~MeV, to the one induced by the mixing of interaction eigenstates), as well as a drop in the couplings (since the vertex $\tilde{H}^0\,\tilde{H}^+\, W^-$ as a coupling which is a factor of $\sqrt{2}$ smaller than for Winos) and hence
a drop in $\alpha$; this has to be compensated by a larger $m_{\chi^0}$.

\subsection{Kinetic decoupling}
\label{kineticdecoupling}

The results we presented in previous Subsections assume that the temperature of the neutralinos traces the thermal bath temperature. This is true as long as neutralinos are in kinetic equilibrium. After kinetic decoupling, at the temperature $T_{\rm{kd}}$, their temperature decreases with the scaling as appropriate for non-relativistic particles, i.e. $T_\chi \sim 1/a^2$, where $a$ is the Universe scale factor, cooling much faster than thermal bath states for which $T\sim 1/a$. This does not have any influence on the relic density computation if interactions do not depend on velocity, as in the standard case of s-wave annihilations. When the Sommerfeld effect plays a major role, however, there is indeed a strong dependence on velocity and colder neutralino may have larger annihilation cross section.  Hence, if the kinetic decoupling happened early enough, i.e. when the depletion term in the Boltzmann equation is still active, this might have given rise to stronger relic density suppression by the Sommerfeld enhancement.

\FIGURE{\label{fig:kd}
	\includegraphics[scale=1.0,bb=-31 -32 229 145]{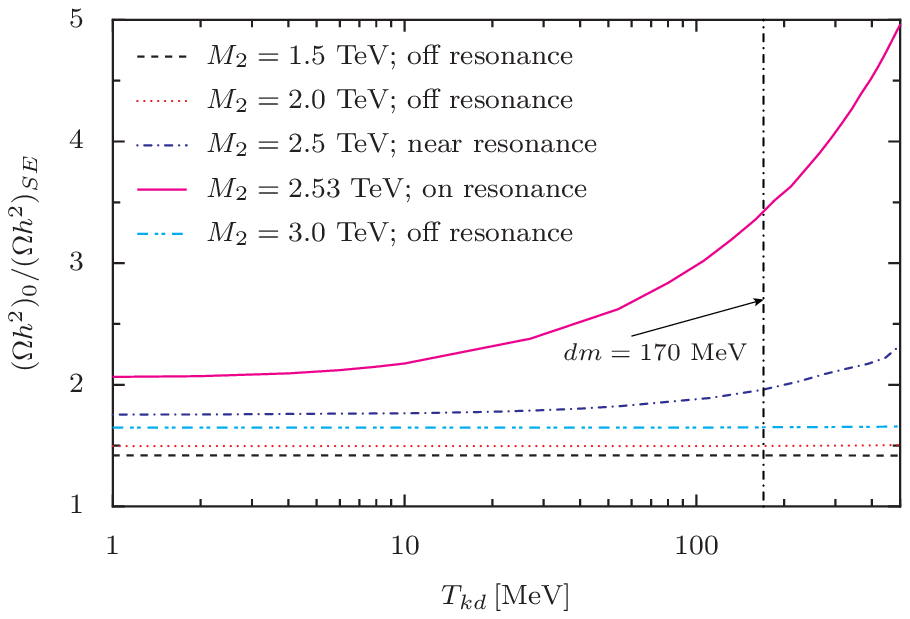}
	\caption{Influence of the kinetic decoupling on the relic density suppression by the Sommerfeld effect for Wino-like neutralinos. In all cases $\mu=5$ TeV and other parameters are as before. For a pure $\tilde W^0$ case the kinetic decoupling occurs at about $T_{\rm{kd}}\approx 170$ MeV, as explained in the text. There is a clear enhancement in the net effect within the resonance region, but no effect away from it.} }

We have just shown that, within the MSSM, the largest impact of the Sommerfeld effect occurs for Wino-like neutralino. In this case, since elastic scatterings are very 
strongly suppressed, the processes enforcing kinetic equilibrium are inelastic scatterings of the type $\chi_1^0 e^- \leftrightarrow \chi^- \nu_e$; the latter are efficient up to the 
time  when the temperature drops down below the mass splitting between neutralino and chargino. Hence, we can estimate that the kinetic decoupling for Winos occurs at about $T_{\rm{kd}}\approx 170$ MeV. The exact value of $T_{\rm{kd}}$ and shape of the distribution function after decoupling (since it's unlikely for the decoupling to be instantaneous) 
depend on the parameters in the model; it is in general rather involved to determine them (see e.g. \cite{Bringmann:2006mu,Bringmann:2009vf}) and this goes beyond the
purpose of this paper. Here, to illustrate the possible effect of the kinetic decoupling on the relic density when the Sommerfeld enhancement is relevant, in Fig.~\ref{fig:kd} we 
plot the ratio between the value of the relic density as computed at the lowest order in perturbation theory $(\Omega h^2)_0$ and the full computation $(\Omega h^2)_{\rm{SE}}$,
for a few points in the parameter space with Wino-like neutralinos, as a function of the value assumed for $T_{\rm{kd}}$ and assuming the neutralino distribution function as 
if decoupling is instantaneous. One can see that in general, the relic density is not sensitive to the kinetic decoupling temperature; the exception is the resonance case, when the 
Sommerfeld enhancement can be much larger and we find a sizable corrections depending on $T_{\rm{kd}}$. It follows that accurate predictions of the relic
abundance in the resonance regime are possible only after determining the kinetic decoupling temperature with a certain accuracy. On the other hand the resonance region
is rather tiny and the overall MSSM picture discussed in this work is not much affected.

\section{Conclusions}

In this paper we have computed the Sommerfeld effect in neutralino and chargino pair annihilations within the MSSM, and discussed its relevance for the thermal relic density of the lightest neutralino. For this purpose, we have adopted and extended the general approach presented in \cite{Iengo:2009ni} to obtain a reliable method of computing the Sommerfeld factors in a case of multi-state systems of fermions with interactions mediated by scalar, vector and axial vector bosons. 

The same method is readily extendable to other models independently of the number of states and interactions that are present; e.g. one could use this same approach within next-to minimal SUSY models (NMSSM), where in some cases a large Sommerfeld effect may be expected due to presence of an additional chiral superfield coupled to Higgs fields.
Moreover, the results presented in this analysis are already readily readable for more general models than the MSSM:  the ingredients we have implemented are only: {\sl i)} the
existence of  massive fermions in the adjoint representation of $SU(2)_L$ (we have been calling them Winos); {\sl ii)} the presence of massive left-handed fermions in the fundamental of  $SU(2)_L$ (without further right-handed ones), which form two doublets with hypercharge assignments such that a $SU(2)_L$ gauge invariant mass is possible (like the $\mu$-term; we have been calling these Higgsinos); {\sl iii)} they form the dark matter and that their $SU(2)_L$ gauge invariant masses are in the range
of TeV; {\sl iv)}  there are $SU(2)_L$ breaking corrections to the mass matrix in the off-diagonal terms in the range of the EW breaking scale, or also in the diagonal terms with a scale comparable to the effect of the off-diagonal ones, such that the resulting masses receive a correction of the order of $O(10^{-2})$ with respect to their
$SU(2)_L$ invariant masses (these determine the mixing among neutral and among charged states; we have been referring to the mass eigenstates from the
superposition of interaction eigenstates as, respectively, neutralinos and charginos). Regarding the last ingredient, we have been computing mass matrices and radiative corrections under the standard MSSM assumptions but this should be just taken as a specific parameter fixing in a more generic model which can be readily generalized.

For our sample numerical study, we have concentrated on neutralino dark matter in the MSSM, being one of the best motivated and most extensively studied scenarios.
Since the Sommerfeld effect is relevant only for non-trivial representations of $SU(2)_L$ we have assumed that the Bino is much heavier and considered only the case of 
Wino--Higgsino mixing; also since in this limit sfermions play a marginal role, we have restricted the analysis to a split SUSY scenario, making the hypothesis that all scalars 
except for the SM-like Higgs are heavy. The computation of the relic density  is based on a numerical solution of the system of coupled Schr\"{o}dinger equations which
allows to compute Sommerfeld factors in all (co-)annihilation channels and then on their implementation in the \ds\ package for a full numerical solution of the Boltzmann 
equation. This enables us to determine the thermal relic density with a very good accuracy, possibly at the few per cent level, in all the parameter space,
except for a small region in which a resonance is found because the Bohr radius becomes of the same order as the interaction range and for which a careful computation of 
the kinetic decoupling process would be needed.

The results we have found show that the Sommerfeld effect can suppress the relic density of the neutralino by as much as a factor of 2-3 in sizable regions of parameter space and even much more in the resonance (the largest suppression factor found in our scan was about 6.2). If the LSP is mostly Wino this results in the shift of the neutralino mass being compatible with WMAP from about 2.2 TeV to about 2.8 TeV, also enlarging slightly the region allowed under standard assumptions on the cosmological model. The existence of the resonance has a small, but visible impact on this regime preferred by cosmology, however it may be much more important if one does not require the neutralino to constitute the whole CDM term. Similar features may occur in other regions of the parameter space; we defer the study of these to future work. 

Qualitatively, the results we have found for pure Winos and Higgsinos are analogous to those in earlier studies of these models, see \cite{Hisano:2006nn,Cirelli:2007xd};
quantitatively, they slightly differ.  There are several reasons for that. 
First of all we were computing the relic density with a full numerical treatment, properly including co-annihilations and Sommerfeld factors, having implemented our new treatment into the \ds\ package. Secondly, we have revised the Sommerfeld enhancement computation itself: we have identified more annihilation channels that need to be treated separately since we have noticed that the enhancement factor depends on the initial spin state of annihilating particle pair, and we have corrected one of the couplings. We also included Higgses exchanges and treated separately two different neutralino states (however, this latter effect is quantitatively not relevant within our parameter space). Moreover, contrary to previous results, our approach allows us to treat not only pure states, but the more realistic situation with arbitrary Higgsino-Wino mixing, with mass spectrum as computed in an arbitrary setup, hence allowing us to explore the full MSSM parameter space.

As a final remark, we would like to point out that, although the numerical work carried out in this paper has been concentrated on the impact on the relic density calculation, the method we developed and the relative numerical routines can be also used to give more accurate predictions for the pair annihilation rate of neutralinos in dark matter halos today
and hence for indirect dark matter detection signals; we will investigate this further in future work.

\acknowledgments

We would like to thank Pasquale Serpico for useful discussions.
The work of A.H. was fully supported and of P.U. was partially  
supported by the EU FP6 "UniverseNet" Research \& Training Network  
under contract MRTN-CT-2006-035863.
A.H. and P.U. would like to thank the Galileo Galilei Institute for  
Theoretical Physics for hospitality during part of the time this work  
was developed. We thank the anonymous referee for providing valuable comments which helped improving the final layout of the paper.

\appendix
\section{Method of computation}
\label{computationmethod}

The form of the $V^\phi_{ij,i'j'}(r)$ for any two-particle states exchanging a boson $\phi$ is always of Yukawa- or Coulomb-type, but due to different couplings and multiplicities we can have different relative coefficients in front. Here we will summarize the way of computing them. Those numerical coefficients depend on the fact what type of fermions, Dirac or Majorana, are present in the diagram and are they distinguishable or not.
\par
In \cite{Iengo:2009ni} it was shown that the integration on the time component of the loop momentum of the Feynman graph expression of the Bethe-Salpeter equation for a four-point amplitude $\Psi$ gives in
non-relativistic limit:
\be
\Psi=V_{int}*{1\ov H_0-{\cal{E}}}\Psi\, .
\label{BS}
\ee
From this expression, by redefining $\Psi\equiv (H_0-{\cal{E}})\Phi$, we get the Schr\"{o}dinger equation
\be
H_0\Phi-V_{int}*\Phi={\cal{E}}\Phi\, .
\label{Sch}
\ee
Here $\Phi$ is in general a multicomponent state describing many possible pairs of particles that 
interact with each other, therefore the equation is just a compact form for a system of equations, and the symbol $*$ states for an operator acting on spins as well.
For each pair $ij$: 
\be
H_0^{ij}=-{\nabla^2\ov 2m_r^{ij}}+2\delta m_{ij}\, ,
\ee
 where $m_r^{ij}=m_i m_j/(m_i+m_j)$ is the reduced mass of the pair and $2\delta m_{ij}=m_i+m_j-(m_a+m_b)$ 
where $m_a$ and $m_b$ are the masses of the incoming particles. ${\cal{E}}$ is the kinetic energy (at infinity) of the incoming pair. 
\par
In order to make explicit  the action of the interaction $V_{int}$ on a state, we express both the interaction and the state in terms of fields
and then make the appropriate contractions.  The relativistic Feynman diagram in the non-relativistic limit gives explicitly the result,
but it is simpler to work directly with the form (\ref{Sch}) by defining suitable non-relativistic contractions. This way of doing the computation has a virtue of easy bookkeeping of all multiplicative factors and signs appearing especially in the case involving Majorana particles. This formalism allows also to include easily the fact that we have initial and final state with defined total angular momentum and spin. 
\par
In the non-relativistic approximation the time delay is neglected and we can work in the time independent Schr\"{o}dinger picture.
\par
We introduce the state $\Phi$ describing a fermion-antifermion (Dirac or Majorana) pair expressed as\footnote{In the following we write $x,y,z,w$ for $\vec x,\vec y,\vec z,\vec w$,  the time coordinate (not indicated) being the same.}
\be
|\Phi^{ij}_\gamma\rangle =N_{ij}\int d\vec{z} d\vec{w}\, \bar\psi_i(z)\mathcal{O}_\gamma\psi_j(w)\ket{0}\Phi^{ij}_\gamma(z,w).
\ee
For a (Dirac) fermion-fermion pair one needs to take $\psi(w)\to\psi^c(w)$. It is easy to see that the spin singlet $S=0$ and the spin triplet $S=1$ are encoded in this formula\footnote{Here we extend the idea presented in \cite{Drees:2009gt} to include also the spin triplet.}:
\be
S=0:~~ \mathcal{O}_\gamma\equiv\gamma_5\, , \qquad S=1:~~ \mathcal{O}_\gamma\equiv\vec\gamma\cdot\vec{S}\, ,
\ee
where $\vec{S}$ is the spin of initial pair. The normalization is 
\bal
&N_{ij}=1/\sqrt{2} \qquad  i\not= j\, , \\ \nonumber
&N_{ij}=1/2~ ~~~~~~~i=j \, .
\end{align}

Take an interaction of the form (as usual by $\Gamma$ we denote the gamma matrices structure)
\be
V_{int}=g_\Gamma^2\int d\vec{x} d\vec{y}\,\bar\psi_k( x)\Gamma\psi_i( x) \bar\psi_j( y)\Gamma\psi_l( y)W_{kl,ij}^\phi( x- y),
\ee
where $W_{kl,ij}^\phi$ is the propagator of the boson exchanged between the two vertices. In the non-relativistic limit only $\Gamma=1,\gamma_0,\gamma_j\gamma_5$ can contribute. The transition $ij\to kl$ can be described in terms of operators acting on the initial state giving the final one: 
\begin{eqnarray}
\label{mastercontr}
&& \frac{g^2_\phi}{2!}\int d\vec{x} d\vec{y}\, \left(\bar\psi_k(x)\Gamma\psi_i(x) +\bar\psi_l(x)\Gamma\psi_j(x)+h.c.\right) \left(\bar\psi_k(y)\Gamma\psi_i(y) +\bar\psi_l(y)\Gamma\psi_j(y) +h.c. \right) W_{kl,ij}^\phi(|\vec{x}-\vec{y}|)\times \cr
&& \times \int d\vec{z} d\vec{w}\, N_{ij} \bar\psi_i(z)\mathcal{O}_\gamma\psi_j(w)\ket{0} \Phi^{ij}_\gamma (z,w)= \cr
&& = \int d\vec{x} d\vec{y}\, N_{kl} \bar\psi_k(x)\mathcal{O}_\gamma\psi_l(y)\ket{0} V_{kl,ij}^\phi(|\vec{x}-\vec{y}|)  \Phi^{ij}_\gamma(x,y).
\end{eqnarray}
The interaction potential between two 2-particle states ($ij\to kl$) arising due to vector, axial vector boson or scalar exchange with the mass $m_\phi$ has the form (see \cite{Iengo:2009ni} for details on the normalization)
\begin{equation}
 V_{kl,ij}^\phi(r)=\frac{c^\Gamma_{kl,ij}}{4\pi}\dfrac{e^{-m_\phi r}}{r}\, .
\end{equation}
We are interested in computing the coefficients $c^\Gamma_{kl,ij(\gamma)}$ for all possible cases.
Note that in the non-relativistic approximation spin-orbit interactions are suppressed, therefore a spin-singlet(triplet) initial state gives a a spin-singlet(triplet) final state, and also the parity of the wave function, that is whether $\Phi(\vec r)=\pm\Phi(-\vec r)$ (e.g. $+$ for the s-wave and $-$ for the p-wave), is the same in the initial and final state. Therefore in general there are four independent systems of equations: the spin-singlet even, spin-singlet odd, spin-triplet even, spin-triplet odd. The coefficients $c_{kl,ij}^\Gamma $ and therefore the interaction potentials are in general 
different in the four cases.
\par
To illustrate the method, consider first a (Dirac) fermion-antifermion pair. We make the possible contractions:
\be
V_{int}*\ket{\Phi^{ij}_\gamma}=g^2 \int _{xyzw}\bar\psi_k( x)\Gamma \langle\psi_i( x)\bar\psi_i(z)\rangle\mathcal{O}_\gamma\langle\psi_j(w) \bar\psi_j( y)\rangle\Gamma\psi_l(y)W^\phi( x- y)\Phi^{ij}_\gamma(z,w).
\nonumber
\ee
By noting that only the creation operator part of both $\bar\psi_a(z)$ and $\psi_b(w)$  appear in the state $\bar\psi_i(z)\mathcal{O}_\gamma\psi_j(w)\ket{0}$,
we get for $p \ll m$:
\bal
&\langle\psi_i( x)\bar\psi_i(z)\rangle\equiv \bra{0}\psi_i( x)\bar\psi_i(z)\ket{0}
=\int {d\vec q\ov 2\omega (2\pi)^3}e^{i\vec p(\vec x-\vec z)} \sum_s u_s\bar u_s 
\to \delta(\vec x-\vec z) P_+\, ,
\\ \nonumber
&\langle\psi_l(w) \bar\psi_l( y)\rangle\equiv -\bra{0}\bar\psi_l(y)\psi_l(\vec w)\ket{0}^T
=-\int {d\vec q\ov 2\omega (2\pi)^3}e^{i\vec p(\vec w-\vec y)} \sum_s  v_s \bar v_s
\to \delta(\vec y-\vec w) P_-\, ,
\end{align}
where $P_{\pm}={1\pm\gamma_0\ov 2}$ and $\omega=\sqrt{\vec{q}^2+m^2}$. Therefore
\be
V_{int}*|\Phi^{ij}_\gamma\rangle =g^2 \int d\vec{x} d\vec{y}\, \bar\psi_k( x)\Gamma P_+\mathcal{O}_\gamma P_-\Gamma\psi_l( y)\ket{0}W_{kl,ij}^\phi( x- y)\Phi^{ij}_\gamma(x,y).
\nonumber
\ee
Note that $\bar \psi_k( x)P_+\ket{0}=\bar\psi_k( x)\ket{0}$ and that $P_-\psi_l( y)\ket{0}=\psi_l( y)\ket{0}$.
\vskip0.2cm
Keeping into account the sign difference between the vector and axial propagator with respect to scalar one and defining $W(\vec r)=e^{-m_\phi r}/4\pi r$, one can write $W_{\gamma_0}=-W,~W_{1,\gamma_j\gamma_5}=+W$. Then by Dirac algebra we finally get 
\be
V_{int}*|\Phi^{ij}_\gamma\rangle =c^\Gamma_\gamma g^2\int d\vec{x} d\vec{y}\, \bar\psi_k( x)\mathcal{O}_\gamma\psi_l( y)\ket{0}W( x- y)\Phi^{ij}_\gamma(x,y),
\ee
with
$$c^1_{\gamma_5}=c^{\gamma_0}_{\gamma_5}=c^{1}_{\gamma_i}=c^{\gamma_0}_{\gamma_i}=1,\qquad c^{\gamma_j\gamma_5}_{\gamma_5}=-3,\qquad c^{\gamma_j\gamma_5}_{\gamma_i}=1.$$
This result gives a term in the equation for $\Phi^{kl}_\gamma$:
\be
H_0^{kl}\Phi^{kl}_\gamma-c_{kl,ij}^\Gamma{e^{-m_\Gamma r}\ov 4\pi r}\Phi^{ij}_\gamma={\cal{E}}\Phi^{kl}_\gamma\, ,
\ee
where $c_{kl,ij}^\Gamma={N^{ij}\ov N^{kl}}c^{\Gamma}_{\gamma} g^2$.
\par
The state may contain Majorana fermions $\chi=\chi^c=C\bar\chi^T$, like 
\be
 \bar\psi(z)\mathcal{O}_\gamma\chi(w)\ket{0}\Phi(z,w),~ \bar\chi(z)\mathcal{O}_\gamma\psi(w)\ket{0}\Phi(z,w),~ \bar\chi_i(z)\mathcal{O}_\gamma\chi_j(w)\ket{0}\Phi(z,w).
 \nonumber
\ee
Again $\mathcal{O}_\gamma=\gamma_5$ for the spin singlet and $\mathcal{O}_\gamma=\gamma_j$ for the spin triplet. Since $\bar\chi(z)$ and $\chi(w)$ contain the creation operator part only, 
the possible contractions between Majorana fermions are, in the non-relativistic limit,
\bal
&\langle \chi( x)\bar\chi( z)\rangle \to \delta(\vec x-\vec z) P_+\, , \qquad \langle \chi( w)\bar\chi( y)\rangle \to \delta(\vec y-\vec w)P_-\, ,
\\ \nonumber
&\langle \chi( x)\chi( w)^T\rangle =\langle \chi( x)\bar\chi( w)\rangle C^T\to -\delta(\vec x-\vec w) P_+C\, , 
\\ \nonumber
&\langle \bar \chi( y)^T\bar\chi( z)\rangle =C^T\langle \chi( y)\bar\chi( z)\rangle \to -\delta(\vec y-\vec z)C P_+\, .
\nonumber
\end{align}
The computations presented above are for the even orbital angular momentum, for instance $l=0$ (s-wave). This means that the 2-body wave function is symmetric. The results for these coefficients are summarized in Table \ref{cTable}. It is very easy to generalize it to higher partial waves. 
\par
These results cover all the possibilities of types of 2-body to 2-body interactions when we can have Dirac or Majorana states, i.e. every other case present in some model will have one of those forms (so the coefficient in Schr\"{o}dinger equation for computing Sommerfeld enhancement will be the same) with possible different coupling constant.

\section{Projection of the (co-)annihilation amplitudes into spin singlet and spin triplet initial states}
\label{projectors}

Since the Sommerfeld effect is in general different for the spin singlet and triplet configurations, it will affect separately the annihilation rates for those two cases. Therefore, we need to compute the relative weight of the two cases. 
\par
We consider the annihilation of two fermionic dark matter particles into two bosonic or fermionic standard model particles. In our approximation we take the initial state to be at rest, and we neglect the dark matter mass differences and the masses of the particles resulting from the annihilation. Hence the kinematics of the annihilation process simplifies; by calling $p_\mu,p'_\mu$ the momenta of the incoming dark matter particles and $k_\mu,k'_\mu$ the momenta of the annihilation products  we get in the CM frame\footnote{It can be done without this simplification but in the case of interest it does not play any role, especially that we are interested only in the relative weights and so by doing this approximation we do not change the value of perturbative annihilation cross section.}:
\be
p_\mu\sim(m,\vec 0)\, ,\qquad p'_\mu\sim (m,\vec 0)\, ,\qquad k_\mu\sim (m,m\vec n)\, , \qquad k'_\mu\sim (m,-m\vec n)\, , \qquad \vec n^2=1\, .
\ee
The propagators of the virtual particles exchanged in the Feynman diagrams representing the amplitude will have denominators 
of the kind:
\be
t-m^2\sim -2m^2\, , \qquad u-m^2\sim -2m^2\, , \qquad s-\mu^2\sim 4m^2\, ,
\ee
where $t\equiv (p-k)^2\sim -m^2, ~~ u\equiv (p-k')^2\sim -m^2, ~~ s\equiv (p+p')^2\sim 4m^2,$ 
and  $m$ is the appropriate mass of the particle mediating the annihilation.
\par
With these simplifications one can decompose the rate to two factors, i.e. rate $\propto K\times \Phi$, where $K$ is the amplitude squared summed over the final spin configurations and averaged over the initial one, and $\Phi$ is the phase space.

The phase space factor, apart from common factors, reads
\begin{equation}
 \Phi=\left[\left(1+\frac{m_1^2}{E^2_{cm}}-\frac{m_2^2}{E^2_{cm}} \right)^2-4\frac{m_1^2}{E^2_{cm}}  \right]^{1/2},
\end{equation}
where $m_1$ and $m_2$ are masses of final particles and $E^2_{cm}$ is the energy in the CM frame.

\begin{figure}[ht]
	\begin{center}
		\label{stdiag}
 		\includegraphics[scale=0.46,bb=71 570 551 720]{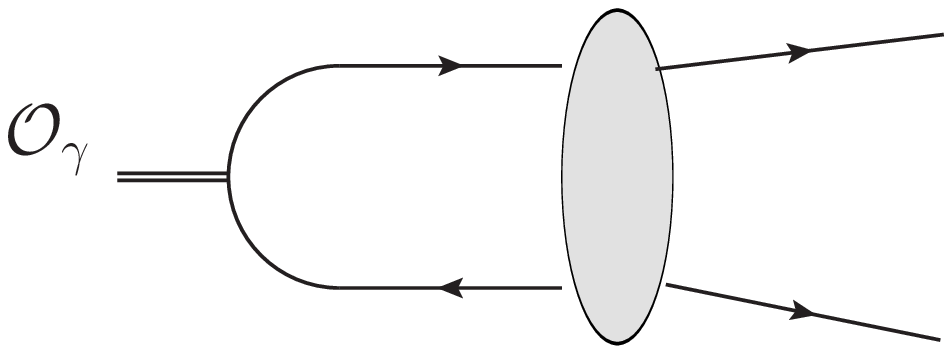}
		\caption{Diagram with explicit contraction of the initial spinors by $\mathcal{O}_\gamma$. The blob represents every possible annihilation process with all possible 2-body final states - fermionic or bosonic.}
	\end{center}
\end{figure}
The relative weights for singlet (triplet) will be computed by the ratio $$\frac{Q(S=0,1)}{Q(S=0)+Q(S=1)}\, ,$$ where $Q(S=0,1)$ is the sum of the singlet (triplet) squared amplitudes for the various annihilation channels:
$$Q(S=0,1)=\sum_{j}Q_{channel~j}(S=0,1).$$
In conclusion, in our approximation, the Sommerfeld enhanced total (i.e. summed over every channel) annihilation cross section $\sigma_{enh}$ is related to the non-enhanced total annihilation cross section $\sigma_0$ by the formula
\be
\sigma_{enh}=\left[{Q(S=0)\ov Q(S=0)+Q(S=1)}S(S=0)+{Q(S=1)\ov Q(S=0)+Q(S=1)}S(S=1)\right]\sigma_0\, ,
\ee
where $S(S=0,1)$ are the Sommerfeld enhancement factors, proportional to the square modulus of the non-relativistic wave function at the origin, for the spin singlet and triplet.

The computation of the weights follows closely the method previously seen. We compute the standard tree-level amplitude of annihilation to given final state but with initial spinors contracted by $\mathcal{O}_\gamma$, as presented with a diagram on a Fig. \ref{stdiag}. In a non-relativistic limit this gives precisely amplitudes $Q(S=0)$ for $\gamma_5$ and  $Q(S=1)$ for $\vec{\gamma}\cdot\vec{S}$. From that point the computations proceed in a standard way.

\bibliographystyle{amsplain}

\begin{thebibliography}{9}


\bibitem{Jarosik:2010iu}
  N.~Jarosik {\it et al.},
  ``Seven-Year Wilkinson Microwave Anisotropy Probe (WMAP) Observations: Sky
  Maps, Systematic Errors, and Basic Results,''
  arXiv:1001.4744 [astro-ph.CO].

\bibitem{Percival:2009xn}
  W.~J.~Percival {\it et al.},
  ``Baryon Acoustic Oscillations in the Sloan Digital Sky Survey Data Release 7
  Galaxy Sample,''
  Mon.\ Not.\ Roy.\ Astron.\ Soc.\  {\bf 401} (2010) 2148
  [arXiv:0907.1660 [astro-ph.CO]].

\bibitem{Riess:2009pu}
  A.~G.~Riess {\it et al.},
  ``A Redetermination of the Hubble Constant with the Hubble Space Telescope
  from a Differential Distance Ladder,''
  Astrophys.\ J.\  {\bf 699} (2009) 539 
  [arXiv:0905.0695 [astro-ph.CO]].

\bibitem{Komatsu:2010fb}
  E.~Komatsu {\it et al.},
  ``Seven-Year Wilkinson Microwave Anisotropy Probe (WMAP) Observations:
  Cosmological Interpretation,''
  arXiv:1001.4538 [astro-ph.CO].

\bibitem{DMbook} \textit{Particle dark matter}, ed. by G. Bertone, Cambridge University Press, 2010.

\bibitem{Adriani:2008zr}
  O.~Adriani {\it et al.}  [PAMELA Collaboration],
  ``An anomalous positron abundance in cosmic rays with energies 1.5-100 GeV,''
  Nature {\bf 458} (2009) 607
  [arXiv:0810.4995 [astro-ph]].

\bibitem{ArkaniHamed:2008qn}
  N.~Arkani-Hamed, D.~P.~Finkbeiner, T.~R.~Slatyer and N.~Weiner,
  ``A Theory of Dark Matter,''
  Phys.\ Rev.\  D {\bf 79} (2009) 015014
  [arXiv:0810.0713 [hep-ph]].


\bibitem{Sommerfeld} A. Sommerfeld, 
 ``\"{U}ber die Beugung und Bremsung der Elektronen'', 
   Annalen der Physik 403, 257 (1931).

\bibitem{Hisano:2002fk}
  J.~Hisano, S.~Matsumoto and M.~M.~Nojiri,
  ``Unitarity and higher-order corrections in neutralino dark matter
  annihilation into two photons,''
  Phys.\ Rev.\  D {\bf 67} (2003) 075014
  [arXiv:hep-ph/0212022].

\bibitem{Hisano:2004ds}
  J.~Hisano, S.~Matsumoto, M.~M.~Nojiri and O.~Saito,
  ``Non-perturbative effect on dark matter annihilation and gamma ray
  signature from galactic center,''
  Phys.\ Rev.\  D {\bf 71} (2005) 063528
  [arXiv:hep-ph/0412403].

\bibitem{Hisano:2006nn}
  J.~Hisano, S.~Matsumoto, M.~Nagai, O.~Saito and M.~Senami,
  ``Non-perturbative Effect on Thermal Relic Abundance of Dark Matter,''
  Phys.\ Lett.\  B {\bf 646} (2007) 34
  [arXiv:hep-ph/0610249].

\bibitem{Cirelli:2007xd}
  M.~Cirelli, A.~Strumia and M.~Tamburini,
  ``Cosmology and Astrophysics of Minimal Dark Matter,''
  Nucl.\ Phys.\  B {\bf 787} (2007) 152
  [arXiv:0706.4071 [hep-ph]].


\bibitem{Cirelli:2008jk}
  M.~Cirelli and A.~Strumia,
  ``Minimal Dark Matter predictions and the PAMELA positron excess,''
  PoS {\bf IDM2008} (2008) 089
  [arXiv:0808.3867 [astro-ph]].

\bibitem{Dent:2009bv}
  J.~B.~Dent, S.~Dutta and R.~J.~Scherrer,
  ``Thermal Relic Abundances of Particles with Velocity-Dependent
  Interactions,''
  Phys.\ Lett.\  B {\bf 687} (2010) 275
  [arXiv:0909.4128 [astro-ph.CO]].

\bibitem{Zavala:2009mi}
  J.~Zavala, M.~Vogelsberger and S.~D.~M.~White,
  ``Relic density and CMB constraints on dark matter annihilation with
  Sommerfeld enhancement,''
  Phys.\ Rev.\  D {\bf 81} (2010) 083502
  [arXiv:0910.5221 [astro-ph.CO]].

\bibitem{Slatyer:2009vg}
  T.~R.~Slatyer,
  ``The Sommerfeld enhancement for dark matter with an excited state,''
  JCAP {\bf 1002} (2010) 028
  [arXiv:0910.5713 [hep-ph]].

\bibitem{Lattanzi:2008qa}
  M.~Lattanzi and J.~I.~Silk,
  ``Can the WIMP annihilation boost factor be boosted by the Sommerfeld
  enhancement?,''
  Phys.\ Rev.\  D {\bf 79} (2009) 083523
  [arXiv:0812.0360 [astro-ph]].

\bibitem{Hannestad:2010zt}
  S.~Hannestad and T.~Tram,
  ``On Sommerfeld enhancement of Dark Matter Annihilation,''
  arXiv:1008.1511 [astro-ph.CO].


\bibitem{Feng:2010zp}
  J.~L.~Feng, M.~Kaplinghat and H.~B.~Yu,
  ``Sommerfeld Enhancements for Thermal Relic Dark Matter,''
  arXiv:1005.4678 [hep-ph].

\bibitem{Iminniyaz:2010hy}
  H.~Iminniyaz and M.~Kakizaki,
  ``Thermal abundance of non-relativistic relics with Sommerfeld enhancement,''
  arXiv:1008.2905 [astro-ph.CO].

\bibitem{Mohanty:2010es}
  S.~Mohanty, S.~Rao and D.~P.~Roy,
  ``Reconciling heavy wino dark matter model with the relic density and PAMELA
  data using Sommerfeld effect,''
  arXiv:1009.5058 [hep-ph].

\bibitem{Arina:2010wv}
  C.~Arina, F.~X.~Josse-Michaux and N.~Sahu,
  ``Constraining Sommerfeld Enhanced Annihilation Cross-sections of Dark Matter
  via Direct Searches,''
  Phys.\ Lett.\  B {\bf 691} (2010) 219
  [arXiv:1004.0645 [hep-ph]].

\bibitem{Buckley:2009in}
  M.~R.~Buckley and P.~J.~Fox,
  ``Dark Matter Self-Interactions and Light Force Carriers,''
  Phys.\ Rev.\  D {\bf 81} (2010) 083522
  [arXiv:0911.3898 [hep-ph]].

\bibitem{Feng:2009hw}
  J.~L.~Feng, M.~Kaplinghat and H.~B.~Yu,
  ``Halo Shape and Relic Density Exclusions of Sommerfeld-Enhanced Dark Matter
  Explanations of Cosmic Ray Excesses,''
  Phys.\ Rev.\ Lett.\  {\bf 104} (2010) 151301
  [arXiv:0911.0422 [hep-ph]].

\bibitem{MarchRussell:2008yu}
  J.~March-Russell, S.~M.~West, D.~Cumberbatch and D.~Hooper,
  ``Heavy Dark Matter Through the Higgs Portal,''
  JHEP {\bf 0807} (2008) 058
  [arXiv:0801.3440 [hep-ph]].

\bibitem{Iengo:2009ni}
  R.~Iengo,
  ``Sommerfeld enhancement: general results from field theory diagrams,''
  JHEP {\bf 0905} (2009) 024
  [arXiv:0902.0688 [hep-ph]].

\bibitem{Cassel:2009wt}
  S.~Cassel,
  ``Sommerfeld factor for arbitrary partial wave processes,''
  J.\ Phys.\ G {\bf 37} (2010) 105009
  [arXiv:0903.5307 [hep-ph]].

\bibitem{Visinelli:2010vg}
  L.~Visinelli and P.~Gondolo,
  ``An integral equation for distorted wave amplitudes,''
  arXiv:1007.2903 [hep-ph].

\bibitem{Gondolo:2004sc}
  P.~Gondolo, J.~Edsjo, P.~Ullio, L.~Bergstrom, M.~Schelke and E.~A.~Baltz,
  ``DarkSUSY: Computing supersymmetric dark matter properties numerically,''
  JCAP {\bf 0407} (2004) 008
  [arXiv:astro-ph/0406204].

\bibitem{Binetruy:1983jf}
  P.~Binetruy, G.~Girardi and P.~Salati,
  ``Constraints On A System Of Two Neutral Fermions From Cosmology,''
  Nucl.\ Phys.\  B {\bf 237} (1984) 285.

\bibitem{Griest:1990kh}
  K.~Griest and D.~Seckel,
  ``Three exceptions in the calculation of relic abundances,''
  Phys.\ Rev.\  D {\bf 43} (1991) 3191.
  
\bibitem{Edsjo:1997bg}
  J.~Edsjo and P.~Gondolo,
  ``Neutralino Relic Density including Coannihilations,''
  Phys.\ Rev.\  D {\bf 56} (1997) 1879
  [arXiv:hep-ph/9704361].

\bibitem{ArkaniHamed:2004fb}
  N.~Arkani-Hamed and S.~Dimopoulos,
  ``Supersymmetric unification without low energy supersymmetry and  signatures
  for fine-tuning at the LHC,''
  JHEP {\bf 0506} (2005) 073
  [arXiv:hep-th/0405159].

\bibitem{Giudice:2004tc}
G.~F.~Giudice and A.~Romanino,
``Split supersymmetry,''
Nucl.\ Phys.\ B {\bf 699} (2004) 65
[arXiv:hep-ph/0406088].

\bibitem{ArkaniHamed:2004yi}
  N.~Arkani-Hamed, S.~Dimopoulos, G.~F.~Giudice and A.~Romanino,
  ``Aspects of split supersymmetry,''
  Nucl.\ Phys.\  B {\bf 709} (2005) 3
  [arXiv:hep-ph/0409232].

\bibitem{Antoniadis:2004dt}
  I.~Antoniadis and S.~Dimopoulos,
  ``Splitting supersymmetry in string theory,''
  Nucl.\ Phys.\  B {\bf 715} (2005) 120
  [arXiv:hep-th/0411032].

\bibitem{Kors:2004hz}
  B.~Kors and P.~Nath,
  ``Hierarchically split supersymmetry with Fayet-Iliopoulos D-terms in  string
  theory,''
  Nucl.\ Phys.\  B {\bf 711} (2005) 112
  [arXiv:hep-th/0411201].
 
\bibitem{Hryczuk:2011tq}
  A.~Hryczuk,
  ``The Sommerfeld enhancement for scalar particles and application to sfermion co-annihilation regions,''
  [arXiv:1102.4295 [hep-ph]].


\bibitem{Pierce:1994ew}
  D.~Pierce and A.~Papadopoulos,
  ``The Complete radiative corrections to the gaugino and Higgsino masses in
  the minimal supersymmetric model,''
  Nucl.\ Phys.\  B {\bf 430} (1994) 278
  [arXiv:hep-ph/9403240].

\bibitem{Cheng:1998hc}
  H.~C.~Cheng, B.~A.~Dobrescu and K.~T.~Matchev,
  ``Generic and chiral extensions of the supersymmetric standard model,''
  Nucl.\ Phys.\  B {\bf 543} (1999) 47
  [arXiv:hep-ph/9811316].


\bibitem{Dine:1992wr}
  M.~Dine, R.~G.~Leigh, P.~Y.~Huet, A.~D.~Linde and D.~A.~Linde,
  ``Towards the theory of the electroweak phase transition,''
  Phys.\ Rev.\  D {\bf 46} (1992) 550
  [arXiv:hep-ph/9203203].

\bibitem{Gross:1980br}
  D.~J.~Gross, R.~D.~Pisarski and L.~G.~Yaffe,
  ``QCD And Instantons At Finite Temperature,''
  Rev.\ Mod.\ Phys.\  {\bf 53} (1981) 43.

\bibitem{Weldon:1982bn}
  H.~A.~Weldon,
  ``Effective Fermion Masses Of Order Gt In High Temperature Gauge Theories
  With Exact Chiral Invariance,''
  Phys.\ Rev.\  D {\bf 26} (1982) 2789.

\bibitem{Press} W.H. Press, B. Flannery, A. Teukolsky, W. Vetterling, \textit{Numerical recipes in FORTRAN 77: The Art of Scientific Computing}, Cambridge University Press 1992

\bibitem{Drees:2009gt}
  M.~Drees, J.~M.~Kim and K.~I.~Nagao,
  ``Potentially Large One-loop Corrections to WIMP Annihilation,''
  Phys.\ Rev.\  D {\bf 81} (2010) 105004
  [arXiv:0911.3795 [hep-ph]].

\bibitem{Bringmann:2006mu}
  T.~Bringmann and S.~Hofmann,
  ``Thermal decoupling of WIMPs from first principles,''
  JCAP {\bf 0407} (2007) 016
  [arXiv:hep-ph/0612238].

\bibitem{Bringmann:2009vf}
  T.~Bringmann,
  ``Particle Models and the Small-Scale Structure of Dark Matter,''
  New J.\ Phys.\  {\bf 11} (2009) 105027
  [arXiv:0903.0189 [astro-ph.CO]].

\end{thebibliography}

\end{document}